\RequirePackage{fix-cm}
\documentclass[pdftex,twocolumn,epjc3,final]{svjour3}  
\smartqed  
\usepackage{xcolor}
\RequirePackage[T1]{fontenc}
\RequirePackage{xcolor}
\RequirePackage{amssymb}
\RequirePackage{amsmath}
\RequirePackage{graphicx} 

\RequirePackage[colorlinks,citecolor=blue,urlcolor=blue,linkcolor=black]{hyperref}
\usepackage{cite}
\usepackage{url}
\usepackage{hyperref,xspace}
\newcommand{\mevnospace}{\ensuremath{{\mathrm{\,Me\kern -0.1em V}}}}
\newcommand{\gevnospace}{\ensuremath{{\mathrm{\,Ge\kern -0.1em V}}}}
\newcommand{\mev}{\mevnospace\xspace}
\newcommand{\gev}{\gevnospace\xspace}
\newcommand{\mbps}{\ensuremath{\text{\,Mbit}/\text{s}}\xspace}
\newcommand{\gbps}{\ensuremath{\text{\,Gbit}/\text{s}}\xspace}
\DeclareMathOperator{\floor}{floor}
\usepackage{comment,mfirstuc}
\journalname{Eur. Phys. J. Plus}

\newcommand{\F}[1]{Fig.~\ref{fig:#1}}

\newcommand{\addComment}[2]{
  \expandafter\newcommand\csname #1\endcsname[1]{{\bf \color{#2} \capitalisewords{#1}:\,##1}}
  \expandafter\newcommand\csname #1cor\endcsname[2]{{\color{#2} \capitalisewords{#1}:\,\st{##1}{\bf ##2}}}
  \expandafter\newcommand\csname #1color\endcsname{#2}
}
\addComment{cris}{blue} 
\addComment{ale}{blue}
\addComment{tom}{red}
\begin{document}

\title{Streaming readout for next generation electron scattering experiments}

\author{Fabrizio Ameli\thanksref{addr_infn_rm1}
        \and
        Marco Battaglieri\thanksref{addr_jlab,addr_infn_ge} 
        \and 
         Vladimir V. Berdnikov\thanksref{addr_cua}
        \and 
        Mariangela Bond\'i\thanksref{addr_infn_ge}
        \and 
        Sergey Boyarinov\thanksref{addr_jlab}
        \and 
        Nathan Brei\thanksref{addr_jlab}        
        \and         
        Laura Cappelli\thanksref{addr_infn_cnaf}
        \and 
        Andrea Celentano\thanksref{addr_infn_ge}
        \and
        Tommaso Chiarusi\thanksref{addr_infn_bo}
        \and    
        Raffaella De Vita\thanksref{addr_infn_ge}
        \and 
        Cristiano Fanelli\thanksref{addr_mit,addr_aifi}
        \and
        Vardan Gyurjyan\thanksref{addr_jlab}
        \and
        David Lawrence\thanksref{addr_jlab}
        \and
        Patrick Moran\thanksref{addr_mit}
        \and
        Paolo Musico\thanksref{addr_infn_ge}
        \and
        Carmelo Pellegrino\thanksref{addr_infn_cnaf}
        \and
        Alessandro Pilloni\thanksref{addr_me,addr_infn_ct}
        \and 
         Ben Raydo\thanksref{addr_jlab}
        \and 
         Carl Timmer\thanksref{addr_jlab}
        \and
        Maurizio Ungaro\thanksref{addr_jlab}
        \and
        Simone Vallarino\thanksref{addr_infn_fe}
}


\institute{
Istituto Nazionale di Fisica Nucleare, Sezione di Roma, 00185 Roma, Italy\label{addr_infn_rm1}
\and
Thomas Jefferson National Accelerator Facility, Newport News, Virginia 23606, USA  \label{addr_jlab}
\and
Istituto Nazionale di Fisica Nucleare, Sezione di Genova, 16146 Genova, Italy\label{addr_infn_ge}
\and
Catholic University of America, Washington, DC 20064, USA\label{addr_cua}
\and
Istituto Nazionale di Fisica Nucleare, CNAF, 40127 Bologna, Italy\label{addr_infn_cnaf}
\and
Istituto Nazionale di Fisica Nucleare, Sezione di Bologna, 40127 Bologna, Italy\label{addr_infn_bo}
\and
Massachusetts Institute of Technology, Cambridge, Massachusetts 02139-4307, USA\label{addr_mit} 
\and
The NSF AI Institute for Artificial Intelligence and Fundamental Interactions, Massachusetts 02139-4307, USA \label{addr_aifi} 
\and
Universit\`a degli Studi di Messina, Dipartimento di Scienze Matematiche e Informatiche, Scienze Fisiche e Scienze della Terra, 98166 Messina, Italy\label{addr_me} 
\and
Istituto Nazionale di Fisica Nucleare, Sezione di Catania, 95123 Catania, Italy \label{addr_infn_ct}
\and
Istituto Nazionale di Fisica Nucleare, Sezione di Ferrara, 44122 Ferrara, Italy\label{addr_infn_fe}
}

\maketitle

\begin{abstract}
Current and future experiments at the high intensity frontier are expected to produce an enormous amount of data that needs to be collected and stored for offline analysis. Thanks to the continuous progress in computing and networking technology, it is now possible to replace the standard `triggered' data acquisition systems with a new, simplified and outperforming scheme. `Streaming readout' (SRO) DAQ aims to replace the hardware-based trigger with a much more powerful and flexible software-based one, that considers the whole detector information for efficient real-time data tagging and selection. Considering the crucial role of DAQ in an experiment, validation with on-field tests is required to demonstrate SRO performance. In this paper we report results of the on-beam validation of the Jefferson Lab SRO framework. We exposed different detectors (PbWO-based electromagnetic calorimeters and a plastic scintillator hodoscope) to the Hall-D electron-positron secondary beam and to the Hall-B production electron beam, with increasingly complex experimental conditions. By comparing the data collected with the SRO system against the traditional  DAQ, we demonstrate that the SRO performs as expected. Furthermore, we provide evidence of its superiority in implementing sophisticated AI-supported algorithms for real-time data analysis and reconstruction.   


\end{abstract}


\vspace{-5mm}

\section{Introduction} 
A new generation of electron scattering experiments is underway at the world-leading QCD facilities such as Brookhaven National Lab (BNL) and Jefferson Lab (JLab).
New projects include the Electron Ion Collider (EIC)~\cite{EIC-YP} at BNL, SOLID~\cite{SOLID} and Moller~\cite{MOELLER} at JLab, and upgrades of the existing detectors in the two labs, sPHENIX~\cite{sPHENIX} and CLAS12~\cite{CLAS12}, respectively.

All these experiments are characterized by modern detectors with millions of active readout channels and by an unprecedented data rate produced by high-luminosity operations of the accelerators.
The ambitious scientific program at the {\it intensity frontier} of nuclear physics calls for a data acquisition system (DAQ) that can record the interesting events and filter out the unnecessary background. Advances in data manipulation  algorithms, e.g. artificial intelligence (AI) and machine learning, open up new possibilities for \mbox{(quasi-)}real-time  data processing, by providing an efficient tool to calibrate the detector while running and at the same time intelligently select and reconstruct the final state particles. To fully exploit this progress, it is necessary to leave the triggered DAQ paradigm and move towards a more flexible software-based framework. In this scheme, all data is streamed from the detector to a data center where the entire detector's information can be analyzed and used for efficient data tagging and filtering.
This framework is called {\it triggerless} or {\it streaming readout (SRO)} DAQ.

\section{Streaming Readout DAQ}
Due to the improved performance of CPUs and computer networks, the FPGA-based trigger scheme can be partially
replaced by SRO DAQ. By removing the hardware trigger and performing the full online data reconstruction, it provides precise selections of (complicated) final states for further high level physics analysis. For example, see the current effort in preparation of the high luminosity upgrade of LHC~\cite{LHCb_allen_2020} at CERN. In a triggerless data acquisition scheme, each channel that exceeds a threshold (as implemented on the front-end board) is labeled with a time-stamp and then transferred, regardless of the status of the other channels. A powerful online CPU farm, connected by a fast network link (usually optical fiber) to the front-end electronics, receives all
data samples, reorganizes the information ordering hits by time, includes calibration constants, and finally applies algorithms to find
specific correlations between reconstructed hits (a.k.a. a software trigger), keeping and storing only filtered events. The advantages of this scheme come from the following: using fully reconstructed hit data to define a high-level event selection condition, implementing online algorithms in a high-level programming language, and gaining the ability to easily upgrade the system configuration and accommodate new requirements. Furthermore, the system can be scaled to match different experimental conditions (either unexpected or foreseen in a planned upgrade) by simply adding more computing (CPUs) and/or data transfer (network switches) resources. FPGAs (or similar fast front-end data concentrators) will still be used in a SRO DAQ system, not to make
decisions concerning which events to select, but to handle ``low-level'' tasks such as adding timestamps or canalizing the data. The pipeline can then be extended to include calibrating and tagging events in the stream. The following will review some examples of how SRO can benefit current and future experiments at electron machines.

\subsection{Real Time Processing: Calibration, Selection, and Tagging}
A significant benefit of SRO systems is the ability to access data from all detectors when making decisions on what must be kept and what can be safely discarded in an event. A natural extension of this concept is to analyze the data as much as is reasonable during acquisition in order to aid and expedite the physics analysis downstream. Two techniques are particularly useful for this. One is the generation of detector calibrations in (quasi-)real time. The other is event tagging. It is with these that SRO starts blurring the line between data acquisition and offline analysis.

\begin{figure}[htb]
\centering
\sidecaption
\includegraphics[width=.48\textwidth,clip]{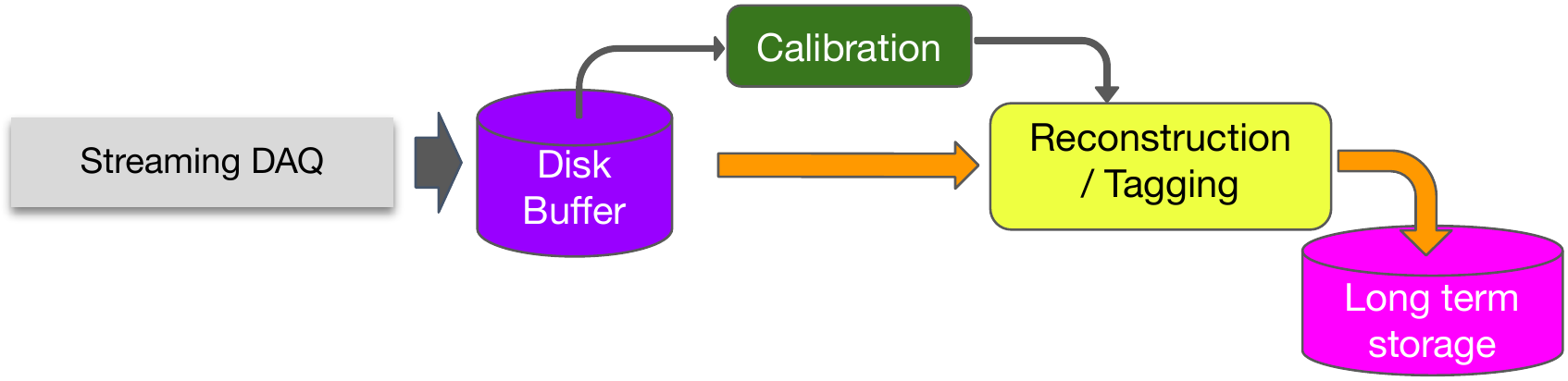}
\caption{Streaming DAQ systems must keep up with the peak rate of the accelerator. However, a large buffer can smooth out fluctuations and downtimes so downstream consumers of the stream need only keep up with the average rate. This gives the added benefit of a time delay that can be used to generate calibrations. Conceptually, this extends the stream all the way into the offline reconstruction.}
\label{fig:2021.10.27.TriDAS_JANA2_EPJA_figures}
\end{figure}

SRO systems must be designed to keep up with the peak detector rates that are determined by the accelerator's full operating luminosity. In actual experiments, accelerators do not operate continuously at full luminosity. Beam trips, fill depletion, configuration changes, etc. all contribute to downtime. Writing the data stream (after triggering and filtering) to a large, temporary disk buffer will smooth out these fluctuations on a time scale corresponding to the size of the buffer. Thus, there is a minimal latency between when data is streamed into the buffer and data is streamed out. This latency can be utilized to obtain detector calibrations from the streamed data before it is sent to storage. Furthermore, with the calibrations available, full or partial reconstruction of the data may be performed prior to storage, further reducing the resource requirements downstream while allowing events to be \emph{tagged} as being of interest to certain physics reactions. Figure~\ref{fig:2021.10.27.TriDAS_JANA2_EPJA_figures} illustrates this concept. The reconstruction/tagging stage shown in figure can be implemented on a large HTC or HPC facility, effectively making it part of the SRO data stream. By way of a quantitative example: existing triggered systems at JLab produce experimental data at peak rates of $1$--$10~\text{GB}/\text{s}$ with approximately a 60\% efficiency of the accelerator. A buffer that gives a 72hr latency would need to be $(10~\text{GB}/\text{s}) \times 0.60 \times (72~\text{hr}) = 1.6~\text{PB}$. The calibration and reconstruction/tagging processes can be done asynchronously with each other while conceptually forming a single, chained stream with multiple data reduction stages.

\subsection{SRO examples}
Here we consider  two specific examples, the new EIC and an upgraded CLAS12 detector at JLab, to discuss benefits of an SRO scheme to the physics program of the two experiments.

\subsubsection{SRO for JLab CLAS12-HI-LUMI}
The CEBAF Large Acceptance Spectrometer for operations at 12 GeV beam energy (CLAS12)  is used to study electro-induced nuclear and hadronic reactions. This spectrometer provides efficient detection of charged and neutral particles over a large fraction of the full solid angle.  Cherenkov counters, time-of-flight scintillators, and
electromagnetic calorimeters provide identification of the scattered electron and of produced hadrons. Fast triggering and high data-acquisition
rates (event rate up to $30~\text{kHz}$, data rate up to $800~\text{MB}/\text{s}$) allow operation at a luminosity of $10^{35}~\text{cm}^{-2}\text{s}^{-1}$. These capabilities are being used in a broad program
to study the structure and interactions of nucleons, nuclei, and mesons, using polarized and unpolarized
electron beam and targets, for beam energies up to 11\gev.
The detector is undergoing to an upgrade aiming to increase the operational  luminosity by a factor of two. In particular the tracking system, currently based on Drift Chambers, will be replaced by a faster tracker with higher granularity such as GEMs, MicroMegas or the novel $\mu$Rwell detectors. The current trigger-based DAQ performance can be increased, reaching an event rate close to $100~\text{kHz}$. For further performance improvement, a full streaming DAQ is under consideration as part of the detector upgrade. The new scheme should overcome other present limitations, such as triggering on neutrals (i.e. photons from $\pi^0$ decay, by imposing a sharp cut on the invariant mass), or triggering on kaons (after performing a crude particle identification).
Improvement is also expected in trigger purity, currently constrained by a significant hadronic background (mainly hadronic showers initiated by pions). More sophisticated algorithms, that make use of the shower distribution and the combined information from the calorimeter and the Cherenkov detector, will result in a cleaner selection of the desired reaction channels. The data rate from the CLAS12 detector in DAQ streaming mode is estimated to be on the level of $50~\text{GB}/\text{s}$, and a data reduction factor 10 or higher is required to decrease the recorded event rate to a level that is compatible with current data storage technology.

\subsubsection{SRO for BNL Electron Ion Collider (EIC)} 
The detectors planned for the future EIC at BNL will be among the few major collider detectors to be built from scratch in the 21st century. A truly modern EIC detector design must be complemented with an integrated readout scheme that supports the scientific opportunities of the machine, improves time-to-analysis, and maximizes the scientific output. A fully SRO
design delivers on these promises.
In particular, the EIC is expected to measure different reactions with an electron (at least) in the electromagnetic calorimeter. Online calibration to compensate for e.m. shower energy leakage and gain variation, implementation of sophisticated AI-supported clustering algorithms for better reconstruction of nearby tracks, and improvement of e.m.-hadron shower discrimination will result in a better resolution, better electron/$\pi^0$ discrimination, and higher hadron background rejection. Moreover, the flexibility of the SRO will allow setting dedicated `triggers' to pin down rare processes, e.g. exclusive kaon-rich reactions, among the leading electromagnetic production. The EIC Yellow Report~\cite{EIC-YP} states that SRO is the chosen option for the collider.

\section{SRO validation and on-beam tests}
Despite the conceptual simplicity of a triggerless DAQ, a realistic implementation with the specific detector readout is necessary to
demonstrate the expected performance. The sophisticated combination of suitable front-end electronics, network facilities, and CPU-based algorithms requires a significant effort to identify or develop (if not available yet) the best option for each element, set up and test the whole scheme, and compare the results with traditional approaches.

In the following, we will describe in detail the components of Jefferson Lab's SRO DAQ prototype (both front-end and back-end) and the results obtained from on-beam tests during an opportunistic measurement campaign performed at the lab in 2020. This prototype is designed to serve as a template for future SRO systems.

\subsection{Detector setup}
On-beam tests were performed with two setups of increasing complexity. In the first experiment, a $3\times 3$ PbWO$_4$ matrix was placed downstream of a secondary electron/positron beam generated by the primary photon beam in Hall-D. Leptons were identified and their energy measured by a Pair Spectrometer tagger (PS)~\cite{Barbosa:2015bga,Somov:2017vhp}. This enabled a precise determination of the energy and a quantitative comparison between SRO and triggered clustering algorithms.
In the second experiment, we measured the inclusive $\pi^0$ electroproduction with the Hall-B CLAS12-FT detector. Detailed Geant4 simulations of the FT provided a realistic estimate of the detector acceptance and efficiency. The $\pi^0$ mass was used as a reference to identify the channel and provide a quantitative comparison to the expected rates. 

\subsubsection{Hall-D}
On-beam tests used an EIC calorimeter prototype consisting of a $3\times 3$ matrix of PbWO$_4$ crystals. Each crystal, $2.05~\text{cm}\times 2.05~\text{cm}\times 20~\text{cm}$ in size (corresponding to $\sim22X_0$) was read out by a $19~\text{mm}$ diameter photomultiplier tube (R4125) with a custom HV base and active divider. 

\begin{figure}
\centering
\includegraphics[width=.4\textwidth,clip]{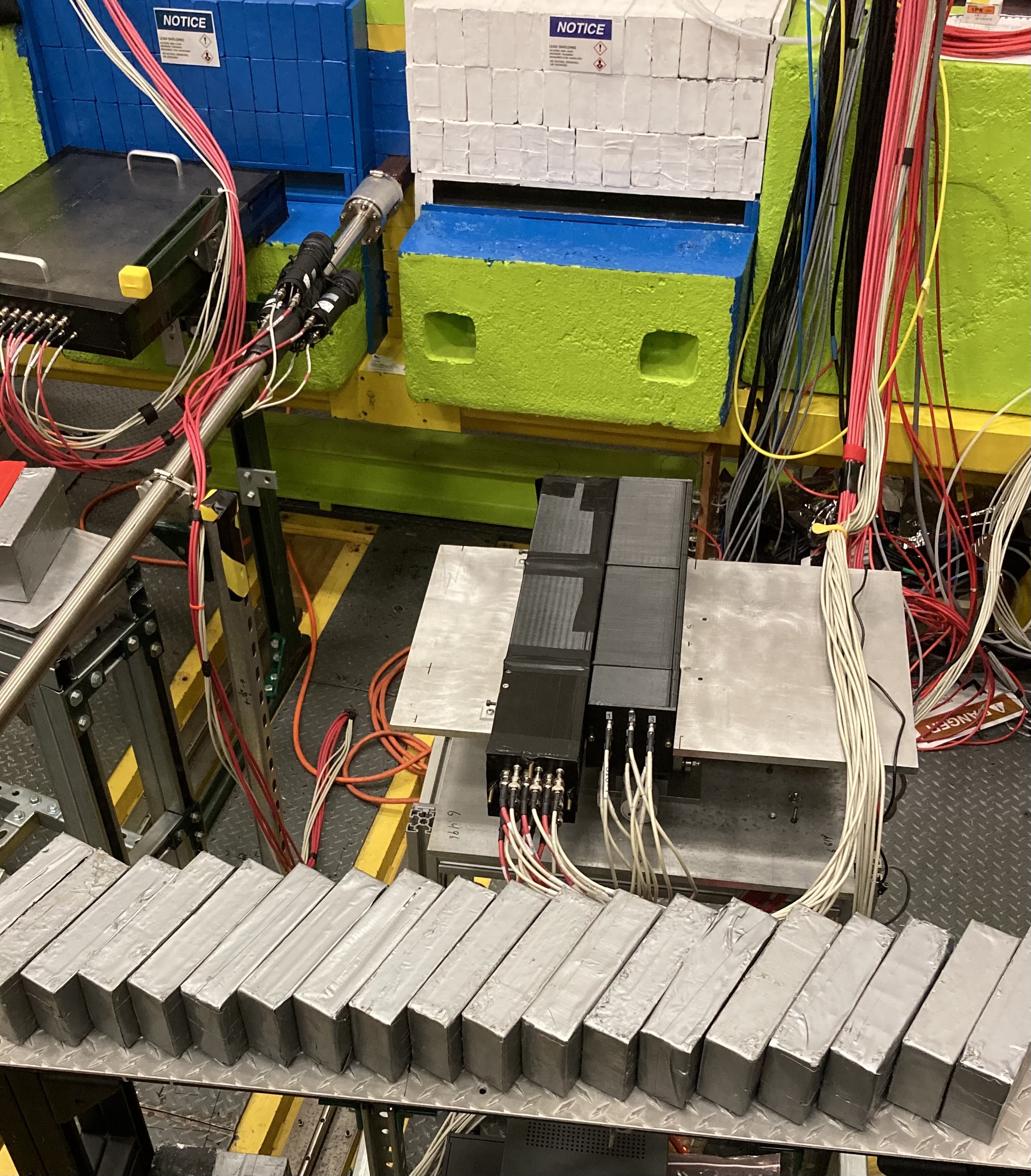}
\caption{The  PbWO$_4$ EIC EM calorimeter prototype (on the left) installed in the Hall-D downstream of the Pair Spectrometer (PS).}
\label{fig:HallD_setup}
\end{figure}

The prototype was installed in Hall-D downstream of the pair spectrometer (PS), as shown in Figure~\ref{fig:HallD_setup}. 
Electron-positron pairs are produced by the primary photon beam interacting with $750~\mu\text{m}$ beryllium converter. Lepton pairs are deflected in a 1.5\,T dipole magnet and detected using two layers of scintillation counters positioned symmetrically around the photon beam line. Each arm consists of 8 coarse counters and 145 high-granularity counters. The high-granularity hodoscope is used to measure the lepton momentum; the position of each counter corresponds to the specific energy. Each detector arm covers the lepton momentum range  $3$--$6.2\gev\!/c$. The energy resolution of the PS is estimated to be better than $0.6\%$. The position of the prototype was surveyed and aligned with respect to the beam line and the center of the pair spectrometer magnet, such that the lepton beam's spot is focused on the center row of the prototype, perpendicular to the front face of the crystals.

\subsubsection{Hall-B}\label{s@hallb}
\begin{figure}[hbt]
\centering	
\includegraphics[width=.40\textwidth,clip]{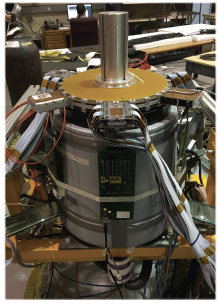}
\caption{The CLAS12 Forward Tagger (FT) with details of the PbWO calorimeter (FT-Cal), the plastic scintillator tile hodoscope (FT-Hodo) and the MicroMega tracker (FT-TRK).}
\label{fig:ft-cal}
\end{figure}

The  JLab SRO DAQ system, which is also expected to be used in upcoming CLAS12 high luminosity operations, was validated on one of the two electromagnetic calorimeters present in CLAS12.
The EM calorimeter provides the trigger for most, if not all, of the processes of interest in electron machines by tagging the scattered electron. In CLAS12 there are two different calorimeters: the FT-Cal  and the FD-Cal.
The FT-Cal covers polar angles (with respect to the beam axis) $2.5^\circ < \theta < 5.0^\circ$, and azimuthal angles $0^\circ < \phi<360^\circ$. The FT-Cal, located  $1.85~\text{m}$ downstream the production target, is made of PbWO$_4$ crystals read by APDs for a total of 332 channels.
The FD-Cal, a sampling calorimeter made of lead and plastic scintillator bars and read by PMTs, covers the region $5.0^\circ < \theta < 35.0^\circ$. It is split into six identical sectors, each covering $\sim60^\circ$ in $\phi$, for a total of $\sim 3000$ channels.\\ 
The limited number of channels, the compactness, and the full coverage of the azimuthal angle range together make the FT-Cal an ideal detector for testing the SRO DAQ. Moreover, the FT-Cal is part of a larger system (the CLAS12 Forward Tagger or FT) consisting of a scintillator-tile hodoscope (FT-Hodo) 
and a MicroMega tracker (FT-TRK) located upstream of the calorimeter. The FT-Hodo, composed of 232 plastic scintillator tails, is used to distinguish neutrals from charged particles and, in turn, identifies photons. The FT-TRK, consisting of two layers of MicroMega detectors, allows a precise determination of the electron coordinates. Figure~\ref{fig:ft-cal} shows the detector with the different components.
The whole FT is a small-scale detector that provides charged/neutral particle detection and identification, representing a simple but complete template of the entire CLAS12. For the aforementioned reasons we focused our tests on the FT-Cal. A detailed description of the detector, including on-beam performance, is reported in Ref.~\cite{ACKER2020163475}.\\

\section{JLab SRO DAQ}
On-beam tests were performed using different combinations of front-end electronic boards (INFN-WaveBoard and JLab fADC250), the CEBAF Online Data Acquisition framework (CODA), the Triggerless Data Acquisition System (TriDAS) (for the back-end) and the high-level analysis framework JANA2. 

\subsection{Front-end electronics}

\subsubsection{The INFN WaveBoard digitizer}
\hyphenation{}

\newcommand{\WR}{White~Rabbit\xspace}
\newcommand{\WVB}{{\it WaveBoard}\xspace}

\begin{figure}[!tbh]
\centering
\includegraphics[width=2.5in]{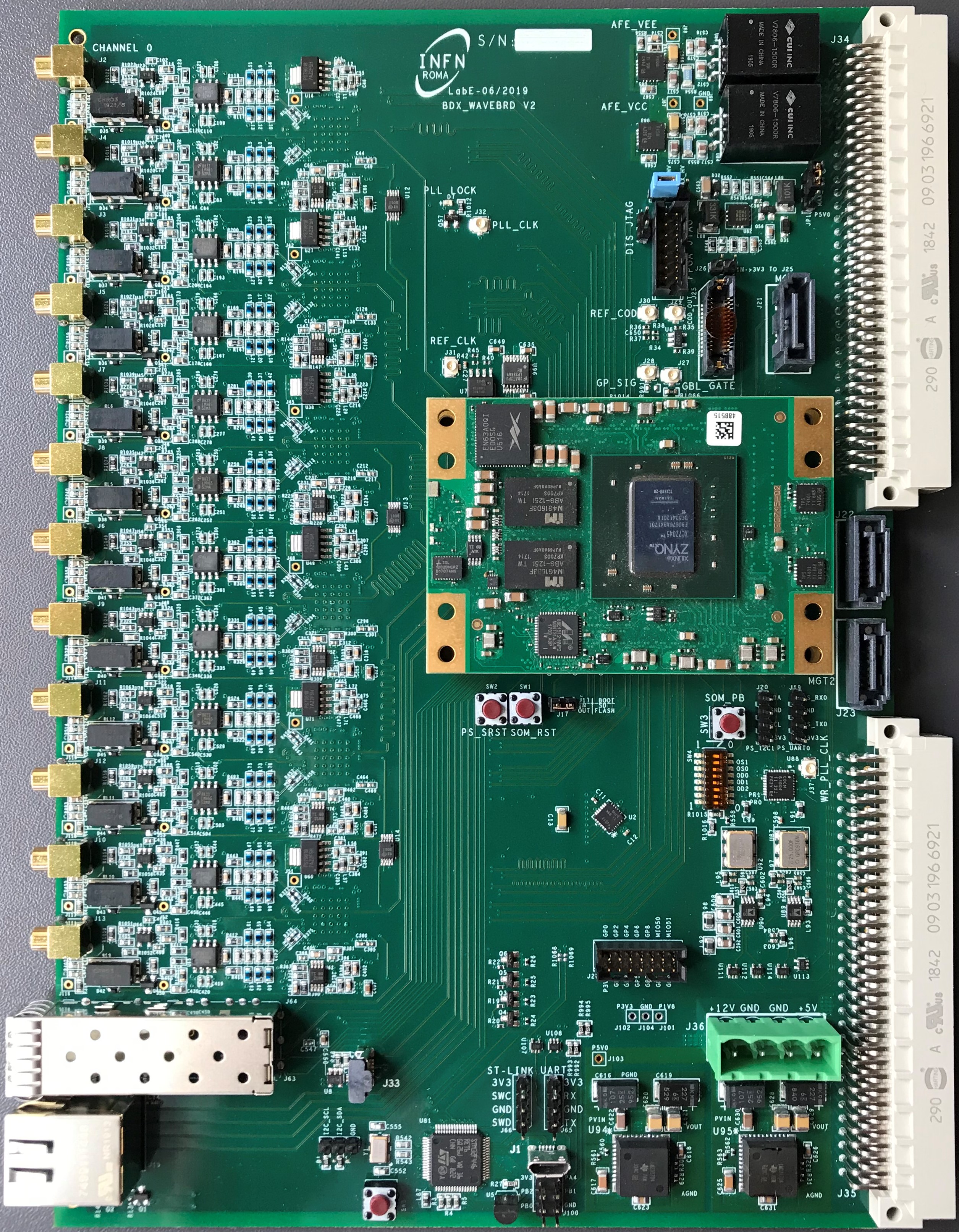}
\caption{The multi-channel digitizer board.}
\label{fig:wvbrd}
\end{figure}

The \WVB is a 12-channel digitizer designed for High Energy Physics experiments. Its main features are an extremely performant sampling architecture at a low cost per channel (compared to equivalent commercial boards), a versatile front-end which can be interfaced to different sensors, a flexible timing system with high resolution, a memory buffer large enough for temporary data storage, and self-triggering algorithms based on waveform analysis.
Fig.~\ref{fig:wvbrd} displays the multi-channel digitizer board. 

The firmware of the digitizer board supports an auto-triggering mechanism particularly suited to the streaming readout environment: digitized waveforms, whose amplitude is greater than a programmable threshold, are stored in a fast FIFO, timestamped, and forwarded to the DAQ via a GbE link for the physics event search. A triggered system can be accommodated as well through dedicated firmware.

Data digitization is performed by 6 dual true differential ADCs from Texas Instruments. The ADC family members are pin to pin compatible, offering resolutions from 12 to 14 bits and sampling frequencies from 65 to $250~\text{MHz}$.

The board is able to power the input sensors individually with a High Voltage (HV) of up to $100~\text{V}$, provided by a dedicated HV linear regulator. 

 To align the system to an external time reference, a clock and a timing signal can be received by the {\WVB} on two dedicated U.FL connectors. The timing signal is fed directly to the FPGA and can be either a slow reference (e.g. a Pulse Per Second signal) or a digital timing protocol (e.g. NMEA, IRIG, etc.). This signal imposes the same phase to timing on different boards.

Data collection and manipulation is accomplished by a commercial System On Module (SoM) mezzanine board from Trenz Electronics, based on a Xilinx Zynq-7000 SoC. The SoM is hosted by the board via proper connectors.
The Zynq-ARM processor runs a Linux distribution, making it substantially simpler to interface with the board. Physics data, stored in DDR memory by the auto-triggering algorithms implemented in the programmable logic, are transmitted over the GbE link using a TCP/IP protocol. Thanks to a Direct Memory Address (DMA) architecture, the available data rate can be as high as $900\mbps$.

\subsubsection{The JLab fADC250 digitizer and VTP board}
\begin{figure}[hbt]
	\centering
	\includegraphics[width=.48\textwidth,clip]{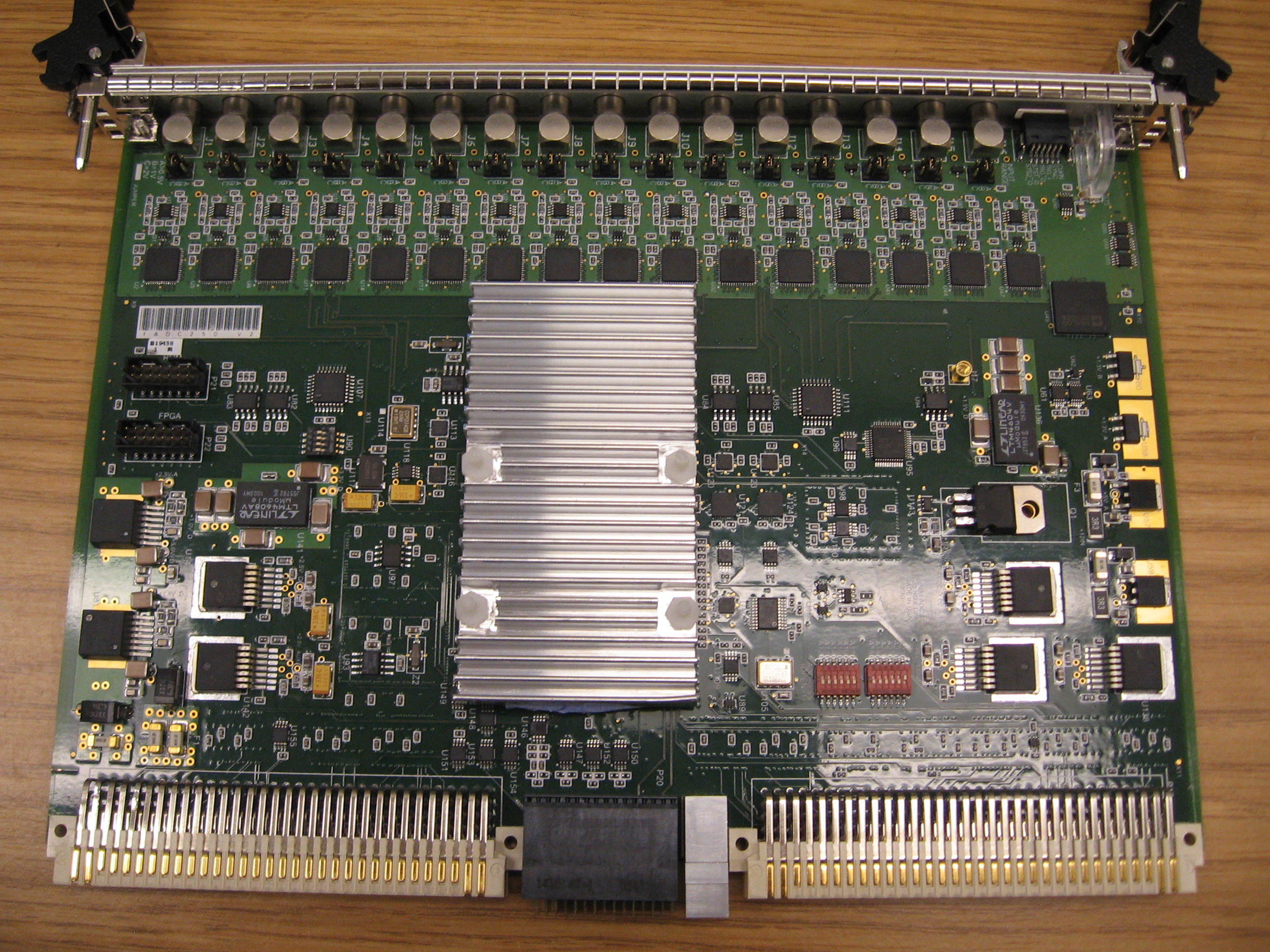}
	\caption{JLab FADC250 Board. A VME/VXS style, 16 channels, $250~\text{MHz}$ 12bit ADCs. Readout can be over VME ($200~\text{MB}/\text{s}$) or through VXS serial links (up to $20\gbps$).}
	\label{fig:fadc250_board}
\end{figure}

The FADC250 (Fig.~\ref{fig:fadc250_board}) is a similar digitizer ($250~\text{MHz}$ sampling rate, 16 channels) designed at JLab and used in many experiments for general purpose triggered readout of detectors. The firmware has been adapted for streaming readout by utilizing the VXS serial links, which were previously used for trigger outputs. The FADC250 supports a total output bandwidth of up to $20\gbps$, though we currently utilize $10\gbps$. The FADC250 firmware detects pulses using a software-defined threshold. When a threshold crossing is found, the pulses are integrated in a programmably-sized window, a pedestal is subtracted, and a gain is applied. The result is a calibrated charge and time for the found pulse, which is sent over the VXS interface to the next stage (VTP). When running in this mode, the FADC250 supports up to $30~\text{MHz}$ of hits per channel.

\begin{figure}[hbt]
	\centering
	\includegraphics[width=.48\textwidth,clip]{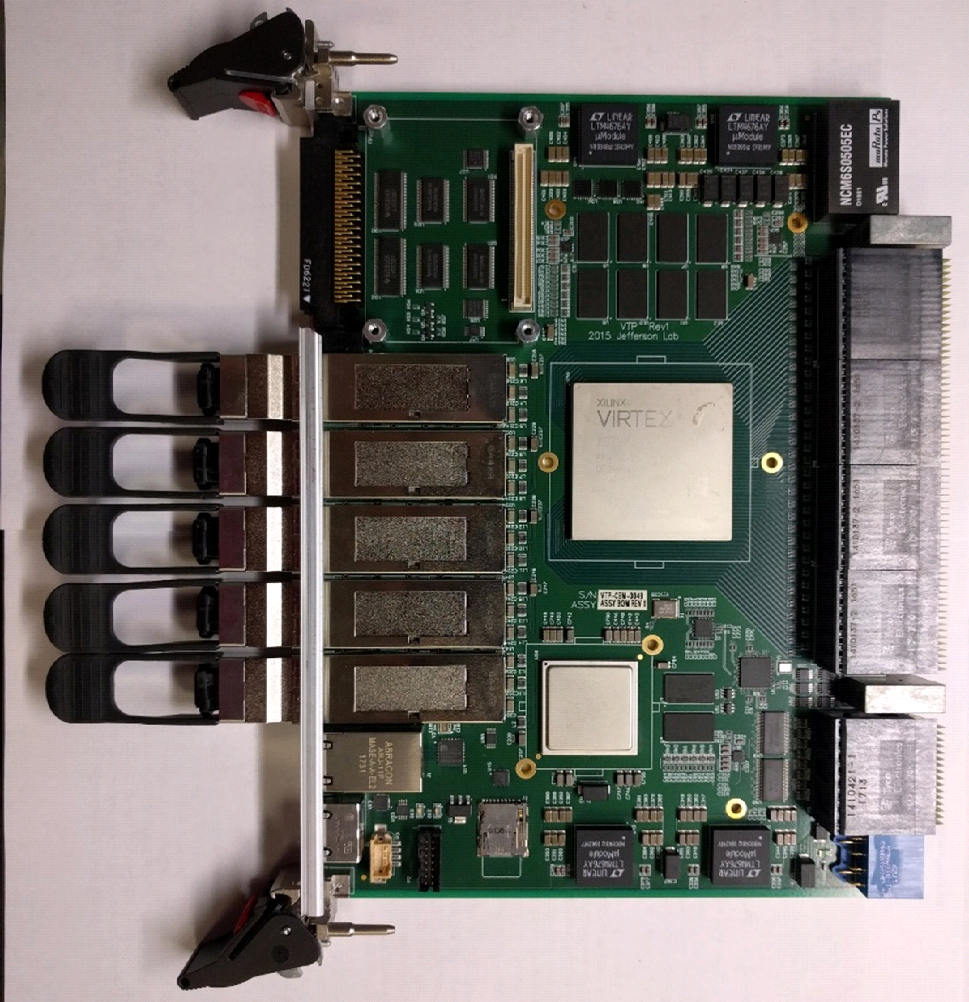}
	\caption{VXS Trigger Processor (VTP) Board. This connects to all VXS front-end cards in a crate to perform readout and/or trigger functions. It provides multiple optical interfaces for sharing information between other VTP modules and for streaming data to the network.}
	\label{fig:vtp_board}
\end{figure}

\begin{figure}[hbt]
	\centering
	\includegraphics[width=.48\textwidth,clip]{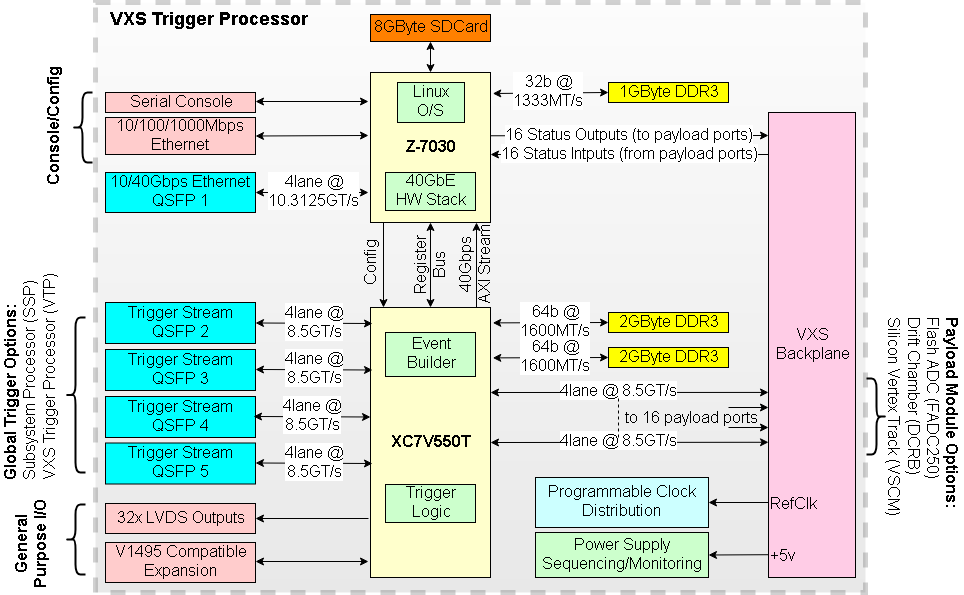}
	\caption{VXS Trigger Processor (VTP) Diagram. Streaming data is collected over VXS from up to 16 payload front-end cards using up to 4 serial lanes at up to $8.5\gbps$ per lane. This data is then buffered in DDR3 memory (connected to the XC7V550T) while the XC7Z030 streams it over Ethernet using up to 4 fully hardware accelerated $10\gbps$ TCP/IP Ethernet interfaces.}
	\label{fig:vtp_diagram}
\end{figure}

The VTP board (Fig.~\ref{fig:vtp_board}) is another custom JLab design that is used in conjunction with FADC250 modules when fast readout is needed. This is a fairly generic and flexible module (with several firmware implementations to match the requirements of different experiments) whose conversion to streaming readout was a simple and natural extension. The resources on the VTP provide  reasonable serial connectivity between the VXS and the optical links, as well as significant buffering capability. See Fig.~\ref{fig:vtp_diagram} for the board schematic.

The VTP collects streaming data from up to 16 FADC250 modules, storing the streams in DDR3 memory. FADC250 hits are packaged into $65~\mu\text{s}$ frames and sent over 10Gbps Ethernet using the TCP protocol. Each frame contains data from up to 8 FADC250 modules. Currently, up to two $10\gbps$ Ethernet links can run together to support up to 16 FADC250 modules. The $10\gbps$ TCP link can sustain up to $8\gbps$ of TCP data without creating significant backpressure. When the data rate is too high and creates TCP backpressure, the streaming data is allowed to back up into the DDR3 memory before any is dropped. If this backpressure is sustained, the VTP will drop data frames. When data is dropped, it will be a complete $65~\mu\text{s}$ frame that is lost; each frame contains a frame counter and timestamp so that the receiving end will know that this happened. To support high data rates without loss, `hot' channels should be distributed over multiple VTPs, and additional $10\gbps$ Ethernet links should be enabled. Currently, when running with 16 FADC250 modules (256 channels) and two $10\gbps$ Ethernet links, the rate limit is around $150~\text{MHz}$ for hits (or about $2~\text{MHz}$ for average hits per channel), which is an extremely large value.

In the future we plan to expand this setup to provide more FADC250 data types and to enable other JLab VXS modules to use the same streaming readout concept, thereby providing a low-cost upgrade opportunity for future JLab experiments.

\subsection{Run Control: CODA}

The CEBAF Online Data Acquisition system (CODA) was designed to work with trigger-based readout systems. The key component is the Event Builder: it collects data from 100+ Readout Controllers (ROCs) and VXS Trigger Boards (VTPs), and builds events based on event number. Another important component is the Trigger Supervisor (TS), which synchronizes all components using clock, sync, trigger and busy signals. ROCs are reading front-end electronics over a VME bus, and VTPs are forming trigger decisions and reporting some trigger-related information. A detailed description of CODA can be found in~\cite{BOYARINOV2020163698}. 

To use the available front-end electronics in streaming mode, the role of the TS was reduced to clock distribution, and the Event Builder was replaced with a new SRO component and back-end software capable of gluing ROC information based on timestamp instead of event number. ROCs are not sending any data in that mode, rather they merely do initial configuration settings over the VME bus. All front-end electronics readout is performed by VTP boards over serial lines rather then the VME bus, which increases the bandwidth limit from about $2$ to $20\gbps$ for each of three participating electronics crates, with the possibility for a further increase to $40~\text{Gbps}$ if needed. New firmware was developed for VTPs to implement streaming mode.

Finally, an intermediate software layer, called {\em SRO Translator}, was introduced between the VTPs and TriDAS. It runs in online mode on its own server. It rearranges and reformats the data streaming from the VTPs so that the data can be ingested by TriDAS. The SRO Translator and TriDAS communicate via TCP sockets over the network. 

\subsection{Triggerless Data Acquisition System (TriDAS)}
\label{TRIDAS_sec}

The Triggerless Data Acquisition System  (TriDAS)~\cite{Chiarusi_2017, TriDAS_JANA2_vCHEP2021} is a streaming readout software framework originally designed and implemented for the astrophysical neutrino detector prototype NEMO~\cite{NEMO}. It is a very large detector, distributed across a cubic kilometer beneath the Mediterranean Sea, and was deployed in stages over several years. A key requirement of TriDAS was to support data taking from the beginning, and to scale with the detector installation. 

To fulfill this ambitious requirement, TriDAS leverages a modular design, which turned out to meet the needs of a beam-based experiment with minimal development effort. TriDAS is made of various software components implemented in {\tt C++11}, each dedicated to a specific task in the data-processing chain.
\begin{figure}[htbp]
    \centering
    \includegraphics[width=.4\textwidth]{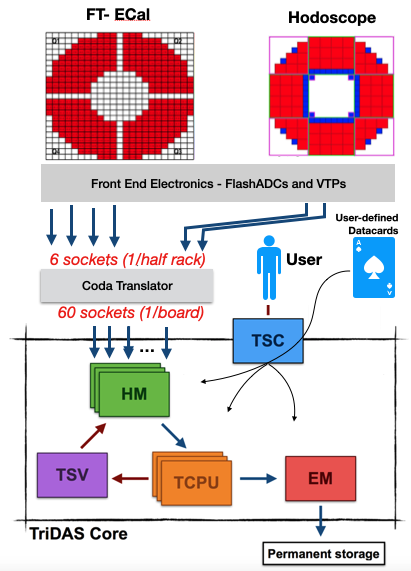}
    \caption{The DAQ model, from the used CLAS12 subdetectors and front-end electronics to the TriDAS elements.}
    \label{fig:TW_corr}
\end{figure}
The HitManagers (HMs) represent the first data aggregation stage. They receive data streams from a predefined number of CODA translators, topologically corresponding to a sector of the detector. Each HM divides the collected data into a sequence of time-ordered bunches of data, called Sector Time Slices (STS), with an adjustable width set to  $50~\text{ms}$. Sharing a common time reference, all HMs arrange their STSs according to the same intervals of time, which are referred to as Time Slices (TS).

The TriggerCPUs (TCPUs) receive the STSes assembled by all HMs corresponding to the same TS and apply the event building and the classification/selection algorithms to the data. Time slices are processed in parallel by multiple threads of the same TCPU and by multiple TCPU processes running on a CPU farm.
The Level~1 events (L1) consist of all data within a programmable time window around a hit whose energy exceeded a certain threshold. Considering the typical sizes of the coincidence time window ($O(100~\text{ns})$) and the TST ($O(10~\text{ms})$), events spanning two adjacent STSes are neglected.
Level~1 events identified within a TS are then fed to the L2 classification/selection algorithms that are implemented in separate binaries and specified in the run configuration file. 

A token-based mechanism is at the core of the TriDAS SuperVisor (TSV) load balancing. Each TCPU thread owns a token that is given to the TSV on completion of the TS processing. The TSV maintains a pool of ``free to use'' TCPU threads which are then matched to the new time slices that are continuously assembled by the HMs.

The Event Manager (EM) collects the selected L2 events and then writes them to the so called Post Trigger (PT) file.

The user interacts with the system via the TriDAS System Controller (TSC). For the CLAS12 tests, a simple interface was custom-built around the TSC.

\subsection{High level data processing (JANA2)}\label{subsec:jana}

One requirement of SRO systems is to be able to filter the data stream so that only the small portion relevant to the physics being measured is ultimately stored. The JANA2 reconstruction framework was integrated with TriDAS for this purpose. 
The solution took the form of a TriDAS plugin which utilizes JANA2. That TriDAS plugin loads multiple JANA2 plugins, each of which provides a different trigger while sharing any underlying algorithms (e.g. calorimeter clustering). Each plugin reports its own \emph{TriggerDecision} back to TriDAS. The \emph{TriggerDecision} contains a $16~\text{bit}$ tag to identify which plugin produced it and a $16~\text{bit}$ value to record details of the trigger decision. Each plugin is free to assign its own meaning to the $16~\text{bit}$ value, with the constraint that nonzero values indicate TriDAS should keep the event. At least one nonzero trigger value is considered sufficient for keeping an event. All \emph{TriggerDecision} objects are stored alongside each recorded event, allowing later analysis of which triggers were active and their specifics for a given event.

\begin{figure}[htb]
\centering
\sidecaption
\includegraphics[width=.48\textwidth,clip]{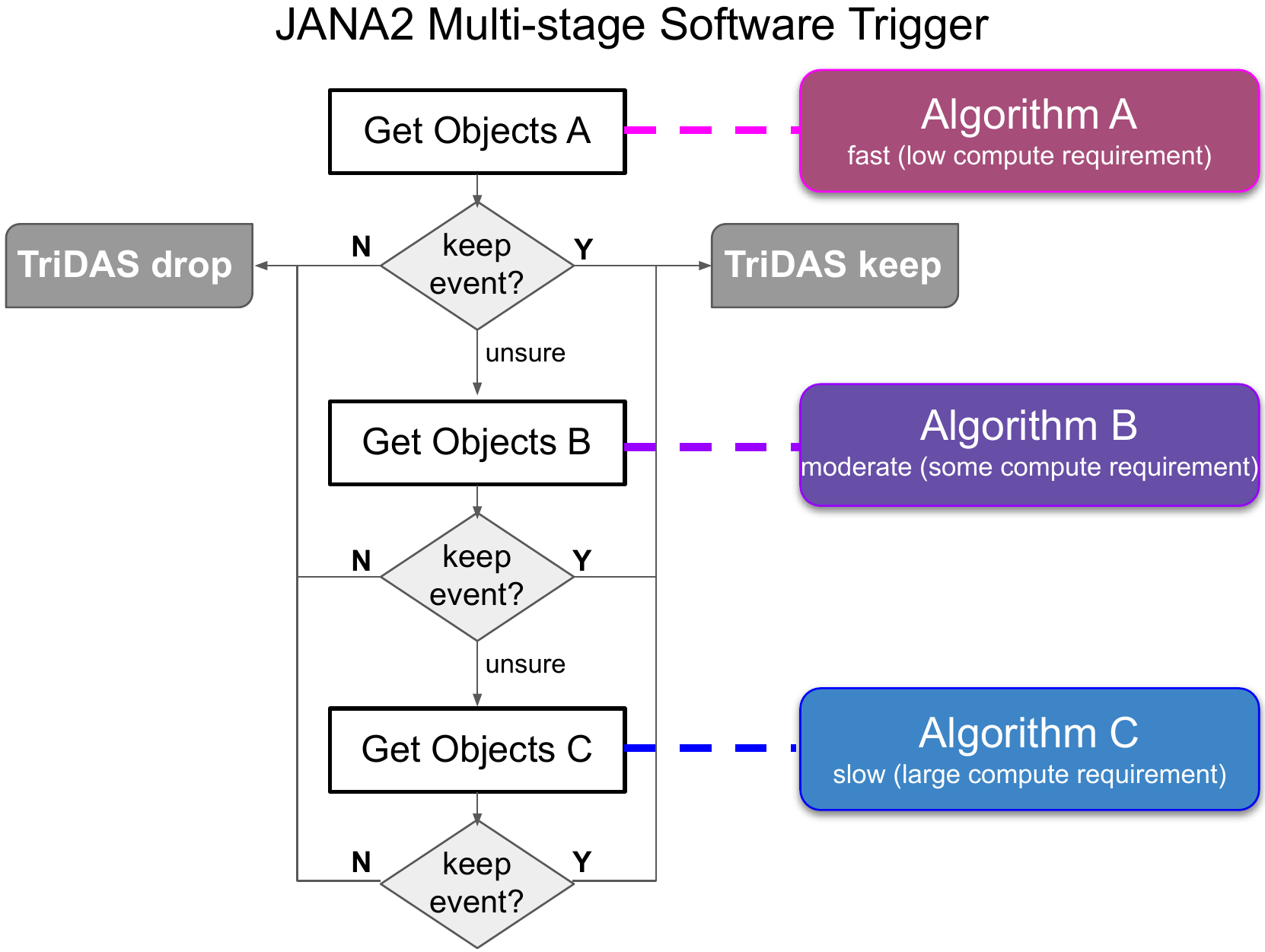}
\caption{Example of a configuration supported in JANA2 where event-by-event decisions are made by applying successively more expensive algorithms. As soon as a keep/no-keep decision is made, TriDAS is informed and additional algorithms need not be run, reducing the overall compute requirement for the online trigger(filter).}
\label{fig:2021.02.11.JANA2_example_trigger_diagram}
\end{figure}

An important design feature of the JANA2 framework is its on-demand algorithm execution. This can significantly reduce the amount of computation required to implement a sophisticated online trigger system. An algorithm in JANA2 (referred to as a ``\emph{factory}'' in the framework's parlance) operates by taking certain data objects as inputs and producing one or more data objects as an output. A user-defined \emph{processor} activates an algorithm by simply requesting the type of data objects that the factory produces. Thus, for a given event, if a certain type of data object is never requested, then the corresponding algorithm that produces it is never run. This allows the user to organize the requests for data objects such that the least expensive algorithms are run first. If a keep/no-keep decision can be made based on their output, no further algorithms need to be run for that event. Figure~\ref{fig:2021.02.11.JANA2_example_trigger_diagram} illustrates a flow chart of this. Having this fine-grained control built-in at the event level helps the end user optimize the system based on the specifics of their data, hardware, and trigger complexity.

\subsection{AI algorithms for SRO}\label{subsec:AI_SRO}

The adoption of AI in nuclear and particle physics is accelerating and will be an essential part of future experiments such as the EIC \cite{AI4EIC_workshop}.

AI encompasses  all  of the  concepts  related  to  the  integration of intelligence into machines; unsupervised learning, in particular, is a family of algorithms capable of learning patterns from untagged data, i.e. without a training phase. This has been explored for the first time in near real-time reconstruction of clusters  detected in the Forward Tagger calorimeter (FT-Cal) using TriDAS and JANA2. 

A clustering algorithm inspired by the \textit{Hierarchical Density-Based Spatial Clustering of Applications with Noise} (HDBSCAN)~\cite{campello2015hierarchical} has been implemented in the form of a plugin within the JANA2 framework.  

 Such an algorithm is characterized by the following features: $(i)$ it can be easily ported to other experiments; $(ii)$ it formally does not depend on cuts during the cluster formation process, making it less sensitive to variations in experimental conditions during data-taking; $(iii)$ it is able to cope with a large number of hits; $(iv)$ it excels when dealing with challenging topologies and arbitrarily shaped clusters, varying cluster sizes and, remarkably, noise (i.e. hits identified as noise are not clustered); $(v)$ it supports calculating the probability that a hit belongs to a cluster or is an outlier. 
 
 The last two features are desirable in many experiments and are not provided by simpler algorithms such as $k$-means~\cite{hartigan1979algorithm}, which is a semi-supervised approach that has also been implemented in the JANA2 framework during these studies.\footnote{In its standard implementation, $k$-means has as hyperparameter the number of iterations to run and the number of clusters $k$; in our implementation of $k$-means, we first determine the seeds of the clusters and then start clustering.} 

HDBSCAN permits considering all of the hit-level information in the detector (e.g. spatial, time, and energy) and looking at the density of hits in that space of parameters. 

\begin{figure}[!] 
\centering
\includegraphics[width=0.48\textwidth]{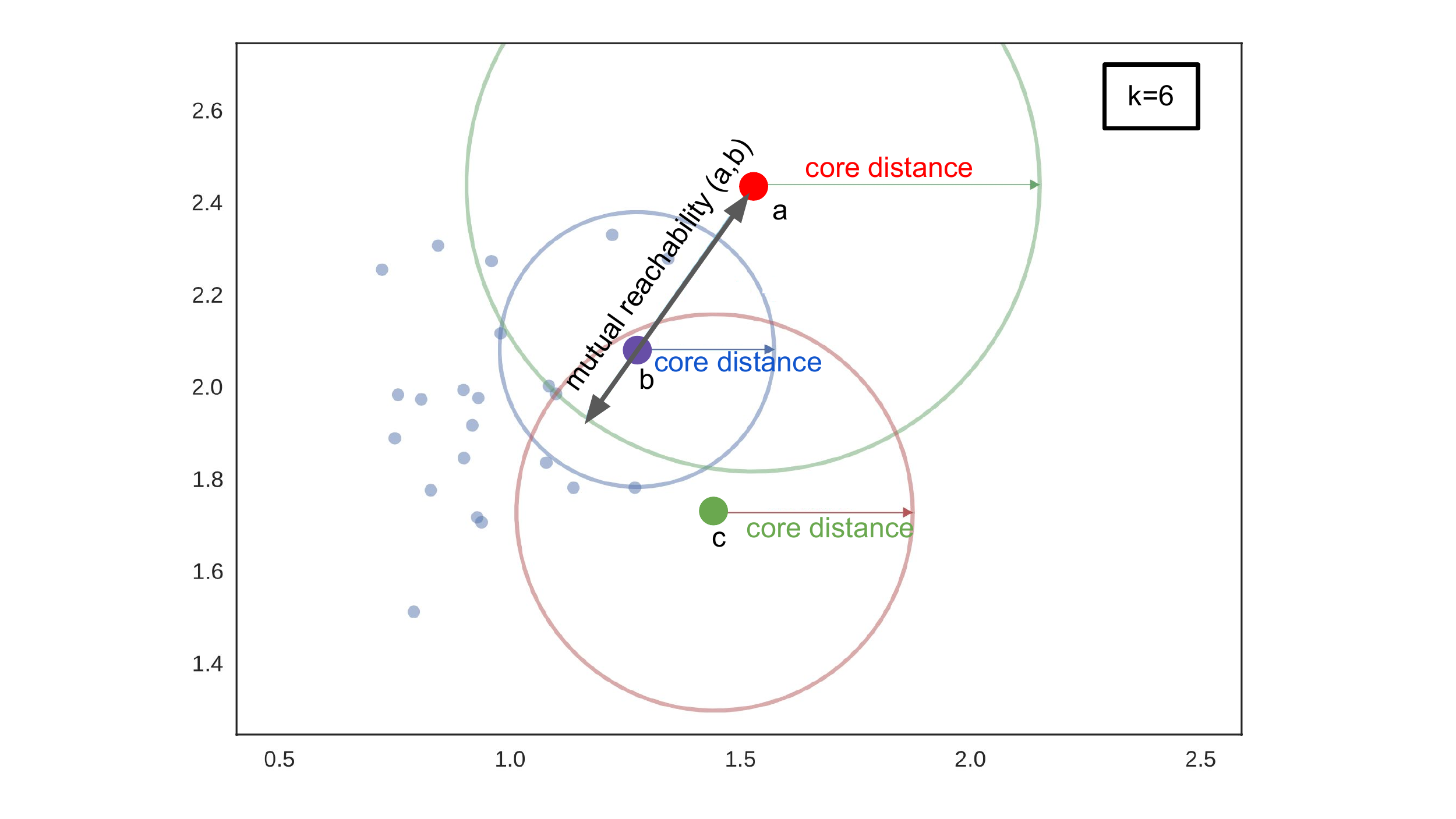} 
\includegraphics[width=0.42\textwidth]{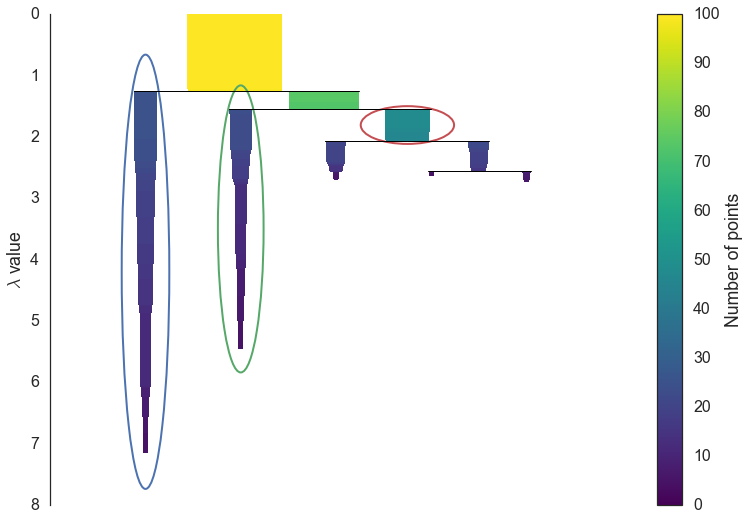} 
\caption{ 
 Illustrative example of mutual reachability (top) for the points $a$ and $b$, following Eq.~\eqref{eq:mutual_reach}. 
 HDBSCAN  condenses the tree upwards via density based notions, defining a maximum lambda boundary (bottom). The horizontal lines indicate the lambda value boundaries of clusters. Image taken from~\cite{HDBSCAN_python}.
\label{fig:ctree} 
}
\end{figure}

To do this, the algorithm leverages a metric called ``mutual reachability'' distance~\cite{campello2015hierarchical}. This metric combines the density at each point (hit) with the relative distance between two points, and is typically expressed as

\begin{equation}\label{eq:mutual_reach}
d_{\text{mreach-}k}(a,b) = \max\left\{\text{core}_{k}(a), \text{core}_{k}(b), d(a,b) \right\},
\end{equation}

where $a$ and $b$ are two points, $d(a,b)$ is their metric distance, and $\text{core}_{k}$ is the core distance at each point, which depends on a hyperparameter $k$, and represents the distance to the $k$th neighboring point. More details can be found in~\cite{HDBSCAN_python}.
This step is followed by finding the minimum spanning tree of the points, based off of the connecting edges for each pair of points with weights related to the mutual reachability. The spanning tree can be converted to a hierarchy which is eventually used to extract the clusters (see Fig.~\ref{fig:ctree}).
In the figure, the parameter $\lambda$ is the inverse of the introduced distance. Another hyperparameter is the minimum cluster size $N_\text{size}$, with which we can now walk through the hierarchy and, at each split, ask if one of the new clusters created by the split has fewer points than the minimum cluster size.
As shown in Fig.~\ref{fig:ctree}, we select clusters from the condensed tree with larger persistence, i.e. with longer lifetime as measured by $\lambda$. 
Recent improvements in the HDBSCAN algorithm can be found in~\cite{hdbscan_imp}. 

This approach is very effective in suppressing single tower noise. 
Hits are associated with clusters along with their probability of membership and their probability of being an outlier.
These become new features that can later be used for refining the selection of physics events. 

\begin{figure}[!] 
\centering
\includegraphics[width=0.44\textwidth]{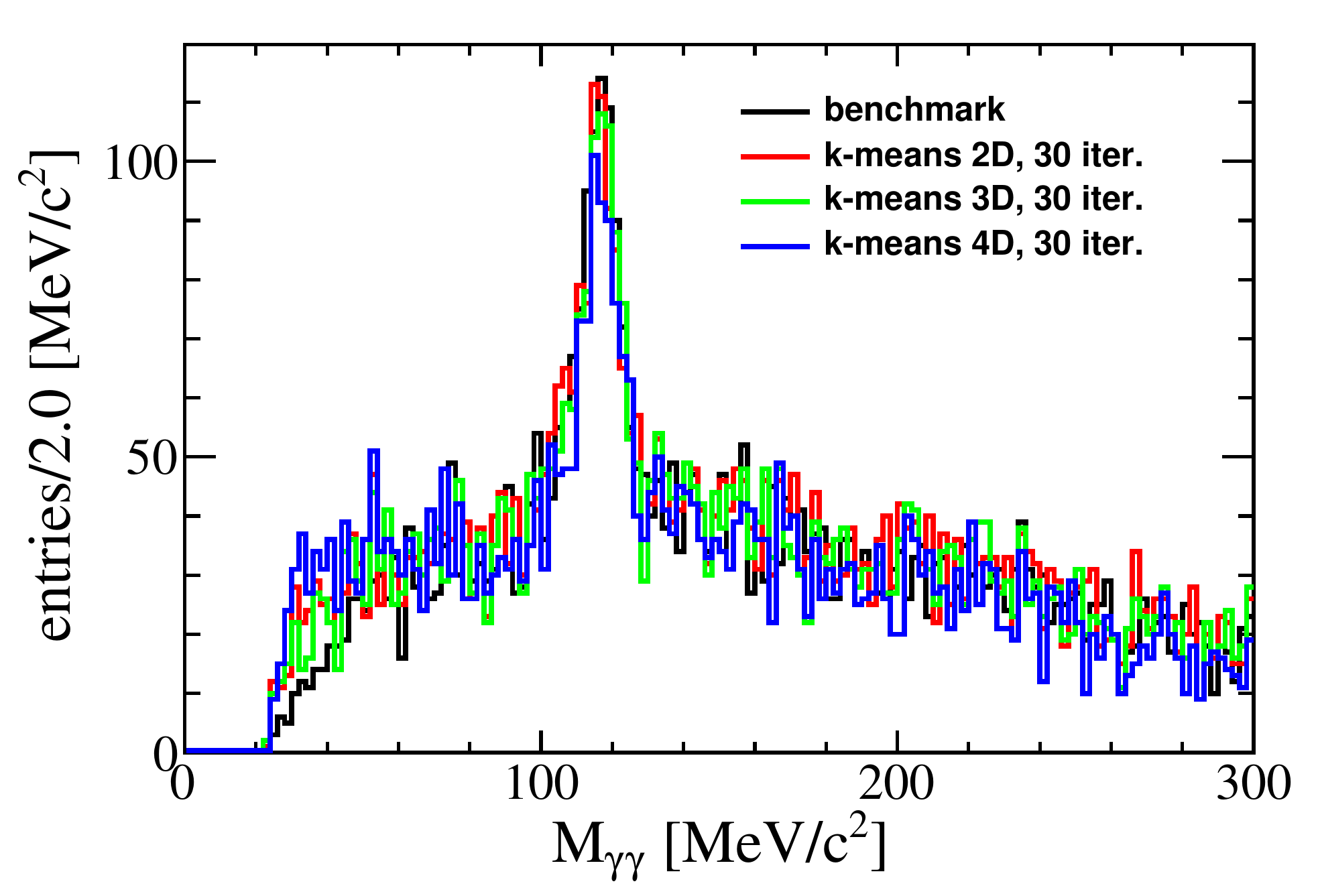} 
\includegraphics[width=0.44\textwidth]{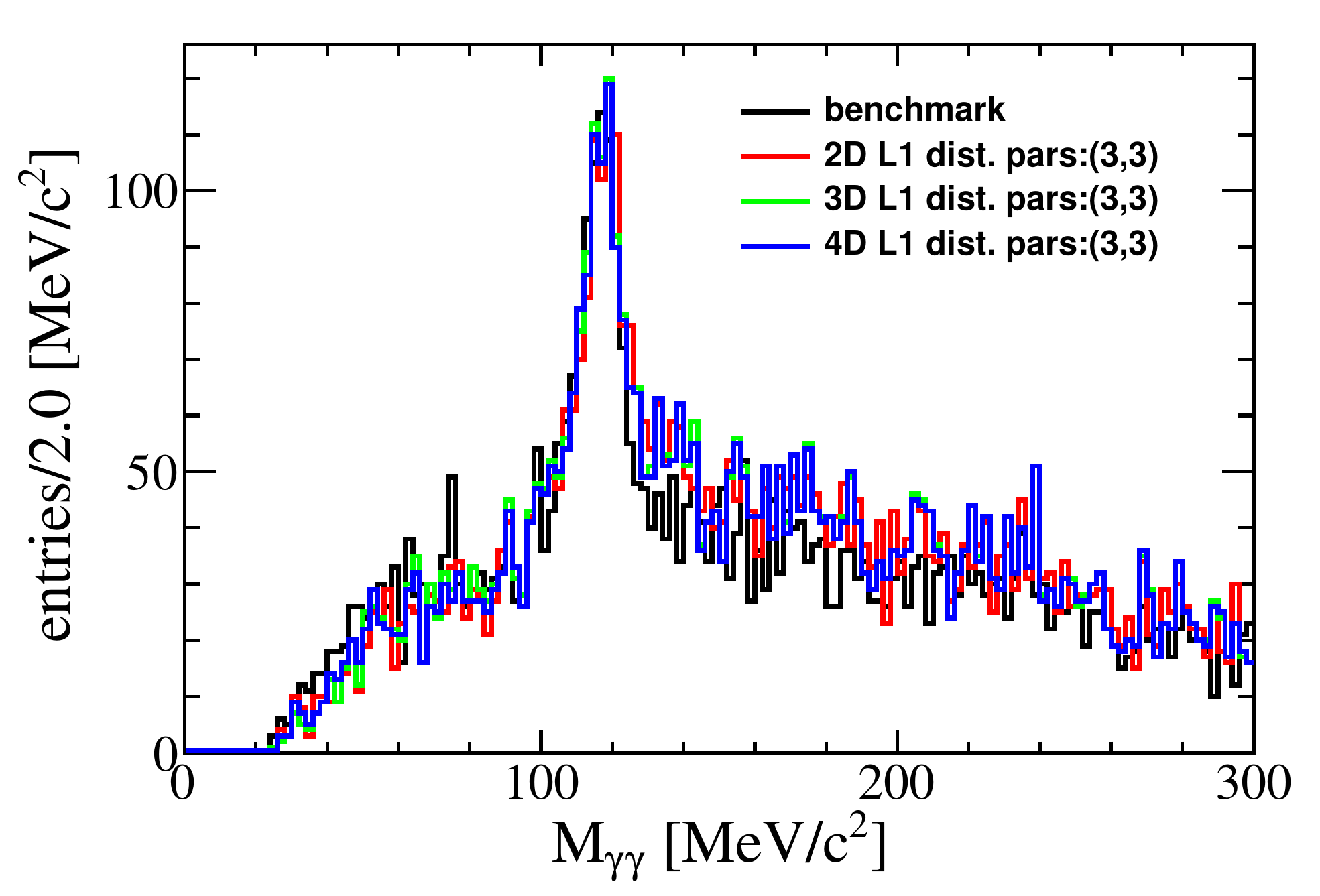} 
\caption{Diphoton invariant mass spectra. Top: standard clustering algorithm (benchmark) and $k$-means, shown with different configurations of the hyperparameters. Bottom: the same  with HDBSCAN.
\label{fig:min_bias} 
}
\end{figure}

Before the SRO tests began, these AI-based approaches were initially developed and tested with well understood and characterized minimum-bias triggered uncalibrated events collected in the FT-CAL of CLAS12, as shown in Fig. \ref{fig:min_bias}.

Diphoton invariant mass spectra obtained with the AI-based approaches have been benchmarked against the spectrum obtained with the standard clustering algorithm to test the tuning of the hyperparameters (Fig. \ref{fig:min_bias} shows only one particular case). 
For HDBSCAN we also extend the clustering to the entire information available in the calorimeter (4D: $x$, $y$, $t$, $E$). Loose selection criteria with fiducial cuts is applied consistently in all cases in Fig. \ref{fig:min_bias} to produce the corresponding diphoton invariant mass spectra. 
With this simple and clean dataset, the $\pi^{0}$ yields obtained with the different methods are comparable, but $k$-means retained more background at lower mass value. As expected, the runtime of $k$-means is comparable to the standard algorithm, while HDBSCAN is 30\% slower on average due to its more complex calculations. 
On the other hand, HDBSCAN is a more suitable clustering strategy for more complex data, as it handles high multiplicity, noise, and complex topologies. 
No cuts on the membership probabilities or outlier scores of the hits have been applied in the HDBSCAN case --- this is a promising opportunity that is left for future studies. 
In Sec.~\ref{sec_data_analysis and results}, we will run the AI-based clustering algorithms on data taken during the SRO tests and provide a detailed description of the accomplished analysis.  

\section{On-beam test results}\label{sec:onbeam_results}
\subsection{Hall-D}

\begin{figure}[hbt]
\centering
\includegraphics[width=.48\textwidth,clip]{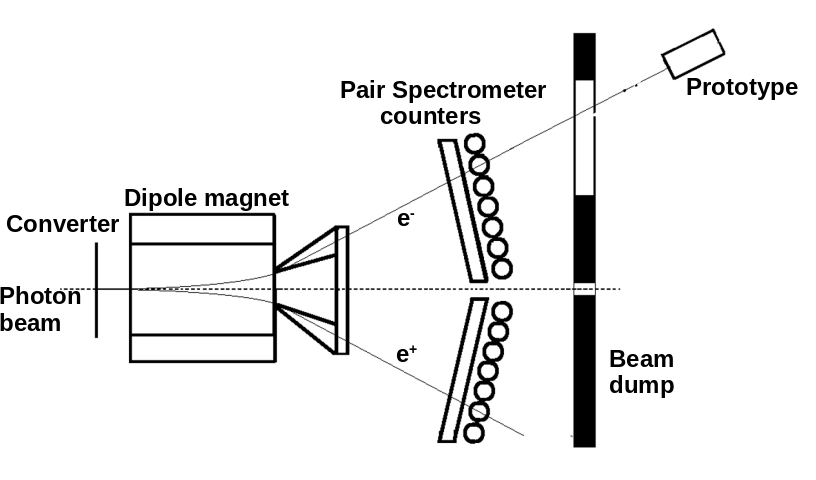}
\caption{Schematic of the prototype tests installed in the Hall-D beamline behind the pair spectrometer}
\label{fig:beamsetup_HallD}
\end{figure}

Tests were performed parasitically during GlueX high-luminosity runs with a $350~\text{nA}$ photon beam.
The prototype was irradiated with a $4.7\gev$ secondary electron beam centered with respect to the matrix central crystal. Figure~\ref{fig:beamsetup_HallD} shows a sketch of the experimental setup.\\
Two different DAQ setups were used: triggered mode (integrated into GlueX data acquisition), and streaming readout. Tests with triggered DAQ were performed by applying the same methodology described in Ref.~\cite{Horn:2019beh}. The signal amplitude from each PMT was recorded by an FADC whenever a lepton hit a PS hodoscope tile.
For SRO tests, each PMT signal was digitized by the {\WVB} and streamed to TRIDAS software, where a threshold equivalent to $\sim2\gev$, defined a L1 event.

\subsubsection{Data analysis and results}

\begin{figure}[hbt]
\centering
\includegraphics[width=.5\textwidth,clip]{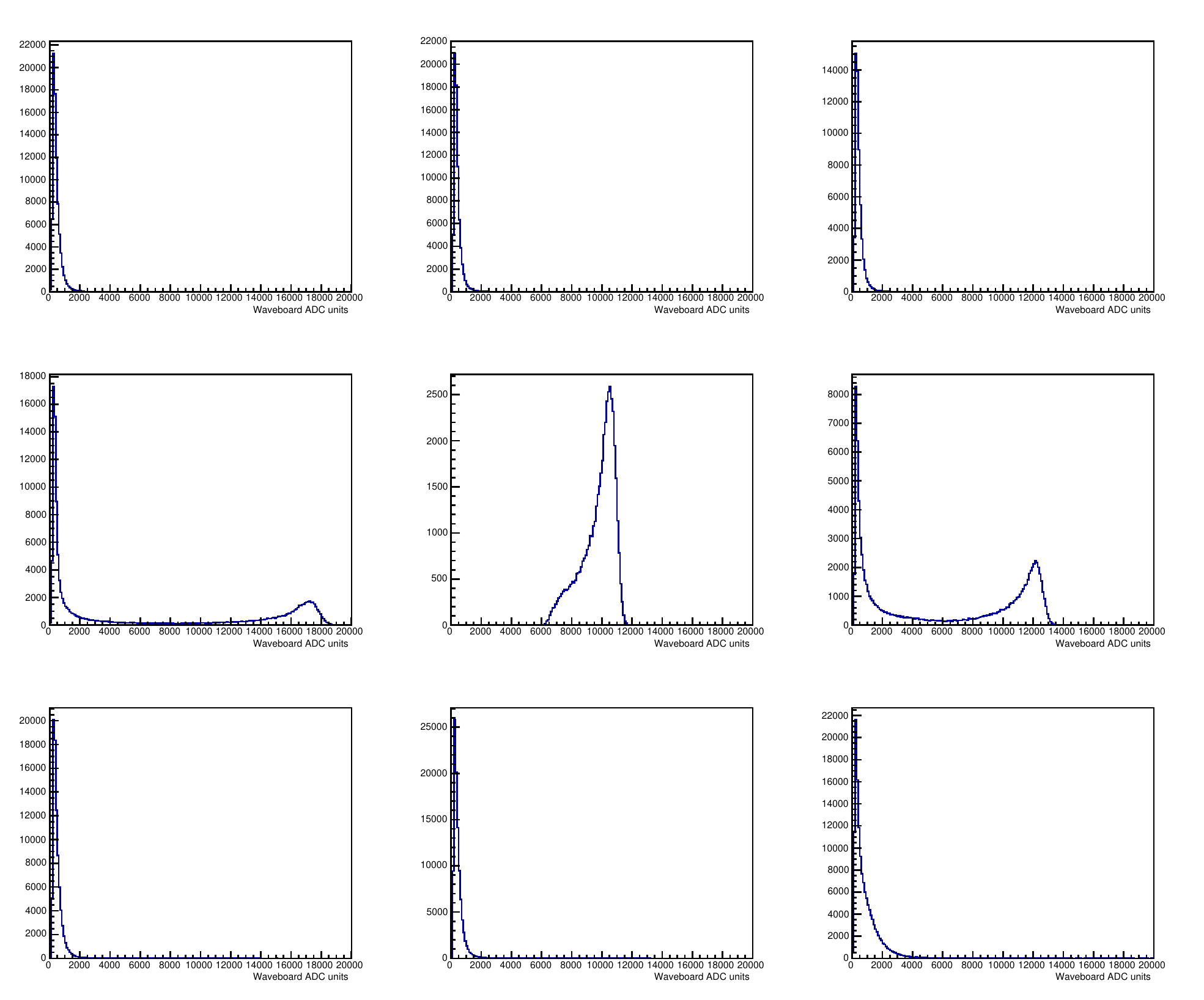}
\caption{Response of the nine channels of the calorimeter collected in  SRO mode. The beam was centered on the central crystal.}
\label{fig:SRO_all_channels}
\end{figure}

To validate the performance of the SRO DAQ chain, we compared the energy resolution obtained in triggered and SRO mode.
The SRO data analysis was performed within the JANA2 framework, where a dedicated clustering algorithm was implemented.
Fig.~\ref{fig:SRO_all_channels} shows the energy spectrum of the nine channels. The effect of the L1 threshold is clearly visible for the central crystal.       

\begin{figure}[hbt]
\centering
\includegraphics[width=.48\textwidth,clip]{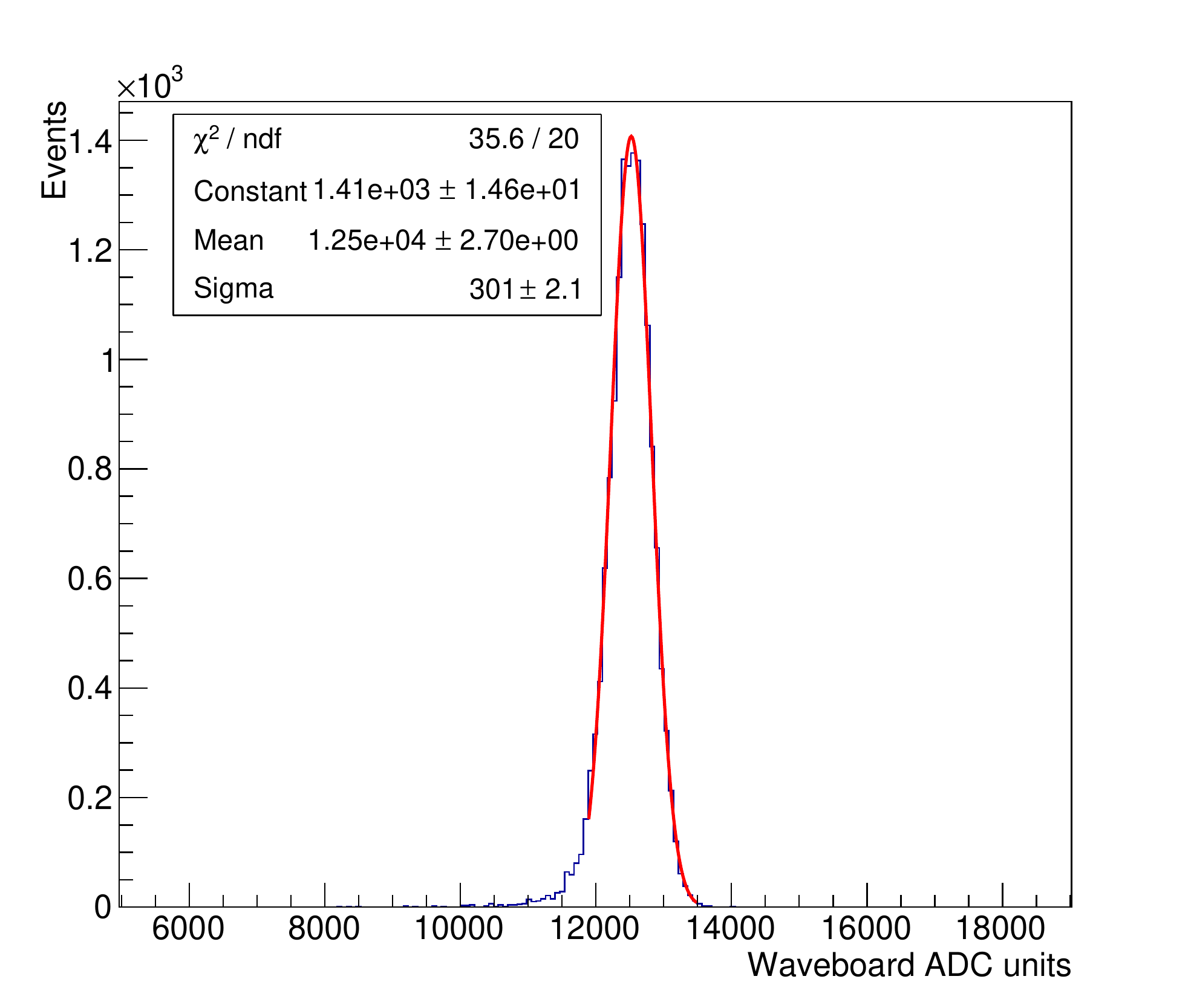}
\caption{Reconstructed cluster energy for the SRO configuration.}	
\label{fig:SRO_Full_cluster}
\end{figure}

The selection algorithm identified events with a large energy deposited in the central crystals (assumed to be the EM shower seed) and summed all hits in the other channels within a time window of $100~\text{ns}$. A cut on the energy-weighted $x$-$y$ hit position was used to exclude events hitting the side crystals after a rough inter-channel energy calibration (the procedure is the same as described later for triggered mode). 
The cluster energy distribution is shown in Fig.~\ref{fig:SRO_Full_cluster}. The energy resolution, obtained by fitting the distribution with a Gaussian curve, is $(2.40\pm0.05)\%$.

\begin{figure}[hbt]
\centering
\includegraphics[width=.48\textwidth,clip]{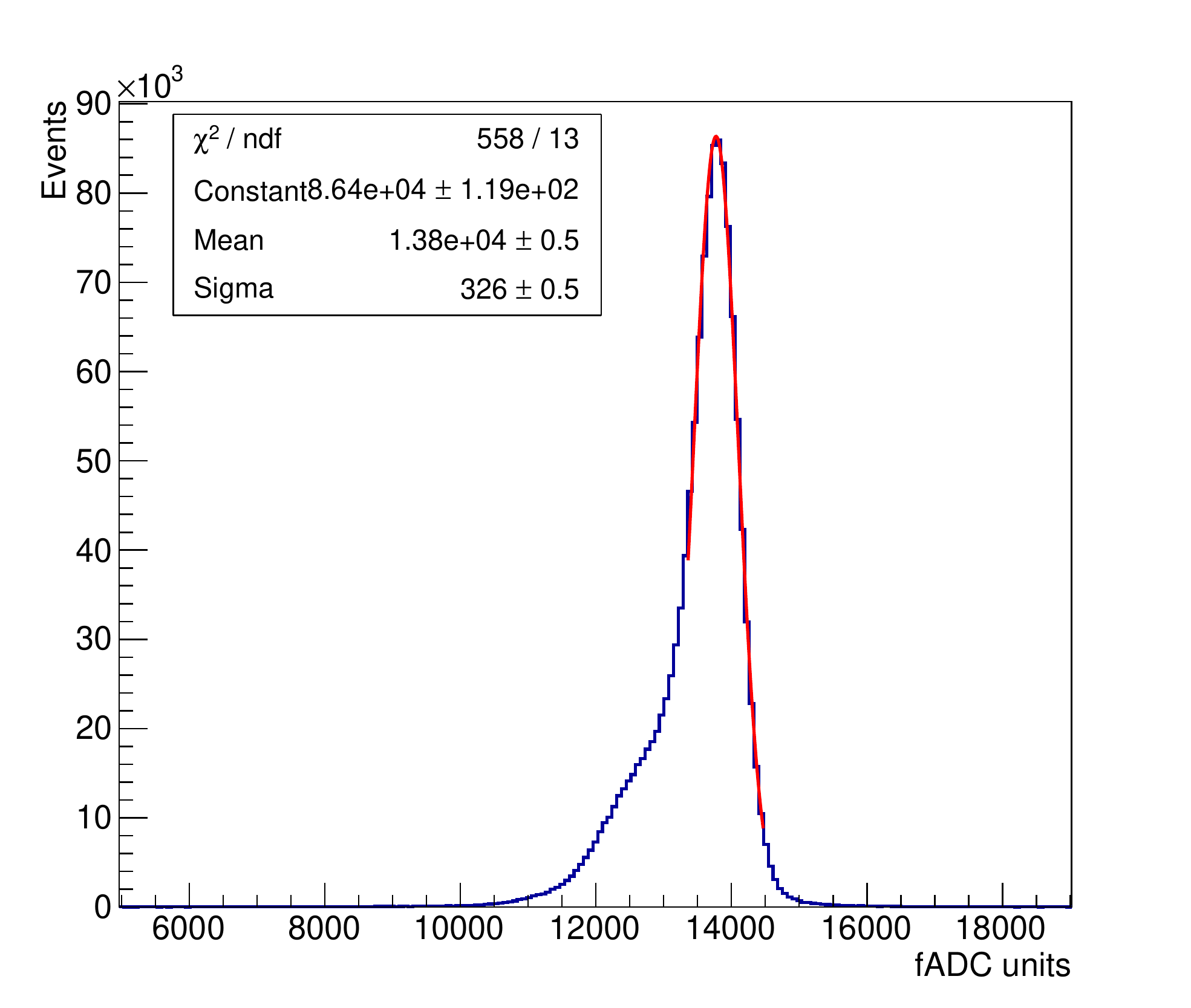}
\caption{Reconstructed cluster energy for the triggered configuration.}
\label{fig:Trig_Full_cluster_wide_cut}
\end{figure}

 The energy resolution in triggered mode was obtained after applying the calibration procedure described in~\cite{Horn:2019beh}. In particular, calibration coefficients were obtained by minimizing the difference between the total energy
deposited in the $3 \times 3$ matrix and the electron energy measured by the Pair Spectrometer.\\
The cluster energy spectrum for the triggered configuration, calculated as the sum of energy deposited in the nine channels, is shown in Fig.~\ref{fig:Trig_Full_cluster_wide_cut}. 

The resulting energy resolution (Gaussian fit)  is $(2.37\pm0.05)\%$,  in good agreement with results obtained for the SRO DAQ configuration. 

\subsection{Hall-B}

Tests were performed parasitically during two CLAS12 production runs using a $100~\text{nA}$ electron beam accelerated at $10\gev$ by the CEBAF on a $125~\mu\text{m}$ lead target for Run-1, and a $40~\text{cm}$ gaseous deuterium target for Run-2.

An SRO system based on FADC250, VTP, TriDAS and JANA2 was used to read out the FT-CAL and FT-HODO.\footnote{During Run-1, only the FT-CAL was used} The TriDAS time coincidence window was set to $200~\text{ns}$ and the TST to $50~\text{ms}$. The energy threshold of the L1 trigger was set to $2\gev$.
Several L2 trigger conditions were used in parallel to tag events: one event over ten, minimum bias, and one or more clusters based on both standard and AI algorithms. 
Minimum-bias triggered data, accumulated in Run-1, was used to identify a clean physics channel consisting of inclusive $\pi^0$ production, and corresponding results were compared with detailed Monte Carlo simulations. In the following sections we report more details about the $\pi^0$ yield estimate and data analysis.

\subsubsection{Inclusive $\pi^0$ electroproduction model} \label{section_theo}
The dominant contribution to the inclusive $\pi^0$ cross section is given by the real and virtual photoproduction of a single $\pi^0$. Real photons can be produced by bremsstrahlung inside the lead target. The expected yield of events is given by~\cite{Tsai:1973py}
\begin{equation}
    Y_\text{real,Pb} = N_e \frac{N_A X_{\text{Pb}0}}{A} \frac{T_\text{Pb}^2}{2} \int_{k_\text{min}}^{E_\text{beam}} \sigma(k)_{\gamma p\to \pi^0p} \frac{dk}{k}\,, 
\end{equation}
where $N_e$ is the number of electrons scattered, $T_\text{Pb}$ 
the thickness of the target in units of radiation length,  $X_{\text{Pb}0}$ the radiation length  times density of lead, and $N_A$ the Avogadro number. For incoherent scattering onto nucleons, $A=1\,\text{g}/\text{mol}$. The cross section is parametrized with the model from~\cite{Mathieu:2015eia}. We consider $k_\text{min}=2\gev$.
The contribution from virtual photons depends on an unknown form factor, function of the photon virtuality. However, it can be approximated by~\cite{Tsai:1973py}
\begin{equation}
    Y_\text{virtual,Pb} \simeq \frac{2 t_{eq}}{T_\text{Pb}} Y_\text{real,Pb}
\end{equation}
with $t_{eq}\simeq 1.7 \times 10^{-2}$. This contribution is about two times larger than the former.

The yields of pions generated by the interaction of the beam with the two exit aluminum foils can be calculated with the same formulae, replacing $\text{Pb}\to \text{Al}$.
Moreover, the target can radiate additional real photons that interact with the lead target. For $80\,\mu\text{m}$ of thickness, this contributes to about 
\begin{align}
    Y_\text{rad,Al} &= N_e \frac{N_A X_{\text{Al}0}}{A} T_\text{Pb} T_\text{Al} \int_{k_\text{min}}^{E_\text{beam}} f(k) \sigma(k) dk\\
    \intertext{with}
    f(k) &=\frac{4}{3}\left(\frac{1}{k} - \frac{1}{E_\text{beam}}\right) + \frac{k}{E_\text{beam}^2} 
\end{align}
with 

We provide the cross sections averaged over the photon energy profile, in a fiducial region $2^\circ < \theta < 6^\circ$:
\begin{subequations}
\begin{align}
    &\frac{1}{\log E_\text{beam}/ k_\text{min}} \int_{k_\text{min}}^{E_\text{beam}} \sigma(k)_\text{fid} \frac{dk}{k} = 182 \,\text{nb} \label{eq:cross1}\\
        &\left[\int_{k_\text{min}}^{E_\text{beam}} f(k)dk\right]^{-1}\int_{k_\text{min}}^{E_\text{b}} f(k)\sigma(k)_\text{fid} dk =  177\,\text{nb} \label{eq:cross1}
\end{align}
\end{subequations}

Subleading contributions can also be considered. The expected largest one is given by $\gamma p \to \pi^0\pi^0 \,p$. As recently measured in~\cite{CLAS:2020ngl}, the $2\pi^0$ invariant mass is dominated by the $f_2(1270)$ resonance. Using the model  from Ref.~\cite{Mathieu:2020zpm}, we estimate that the yields of events having at least one of the pions in the FT acceptance is one order of magnitude smaller than the single $\pi^0$ events, and can be safely neglected.

\subsubsection{CLAS12-FT simulations}\label{subsubsec:simulations}
To provide a realistic estimate of expected $\pi^{0}$ yield, the detector acceptance and reconstruction efficiency was evaluated by  detailed Monte Carlo simulations. 
The geometry, materials and the detector response were simulated using GEMC,  the CLAS12 Geant4 Monte Carlo package~\cite{UNGARO2020163422}.

The geometry was implemented using a database of Geant4 volumes and imported
from the CAD engineering model.

The target volumes implemented in the simulations, shown in \F{ftGeometry}(top) are:
\begin{itemize}
\item a foam scattering chamber with a 50 $\mu$m aluminum window;
\item a $29~\text{cm}$ long cell containing liquid Helium;
\item the lead target, $110~\mu\text{m}$ thick.
\end{itemize}
The FT consists of three subsystems:
\begin{itemize}
\item a tracker (FT-Trk), composed of 4 MicroMega layers;
\item a hodoscope (FT-Hodo), with eight sectors, each containing two layers of scintillators;
\item a calorimeter (FT-Cal) containing an array of 332 crystals.
\end{itemize}
The three subsystems are shown in \F{ftGeometry}-Bottom.
\begin{figure}
\centering
\includegraphics[width=0.99\columnwidth,keepaspectratio]{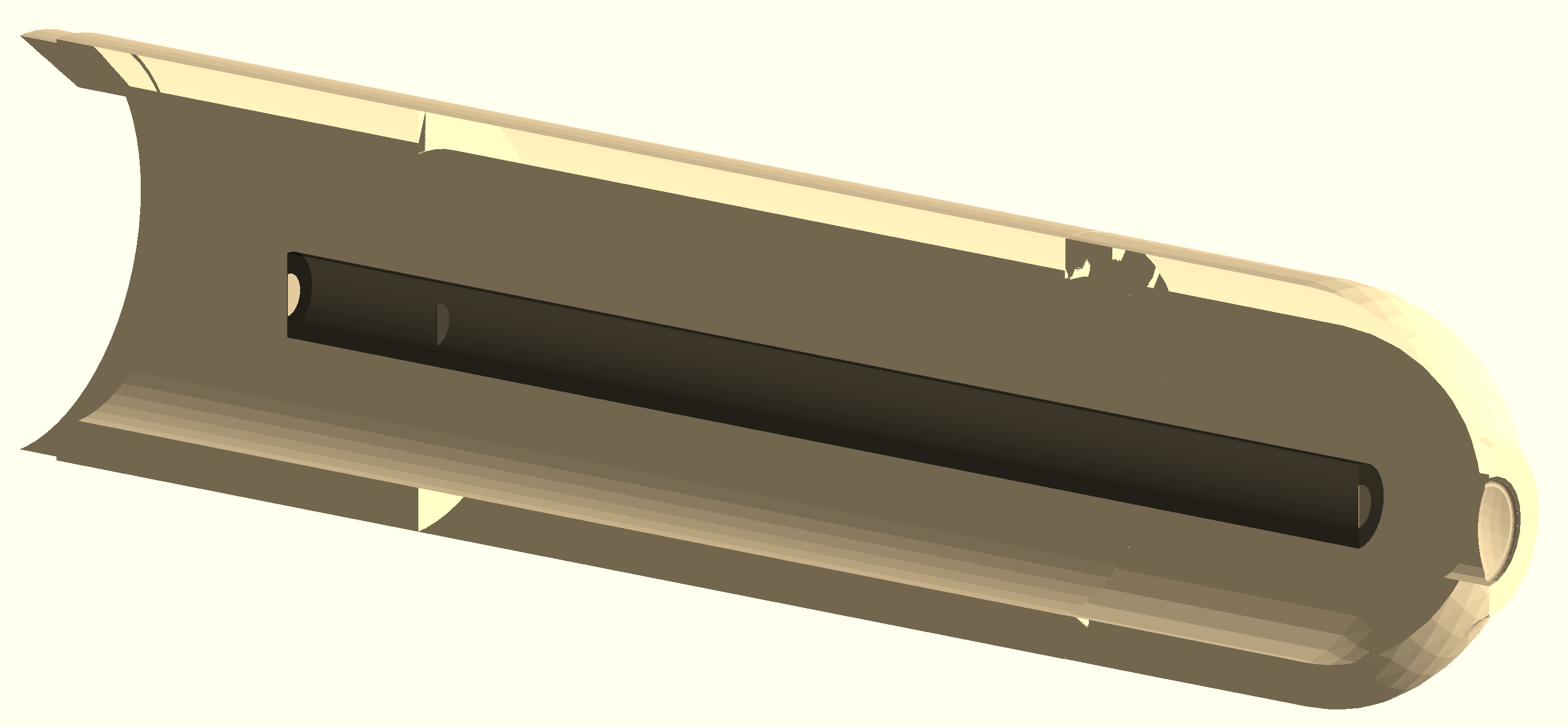}
\includegraphics[width=0.99\columnwidth,keepaspectratio]{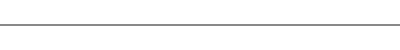}
\includegraphics[width=0.99\columnwidth,keepaspectratio]{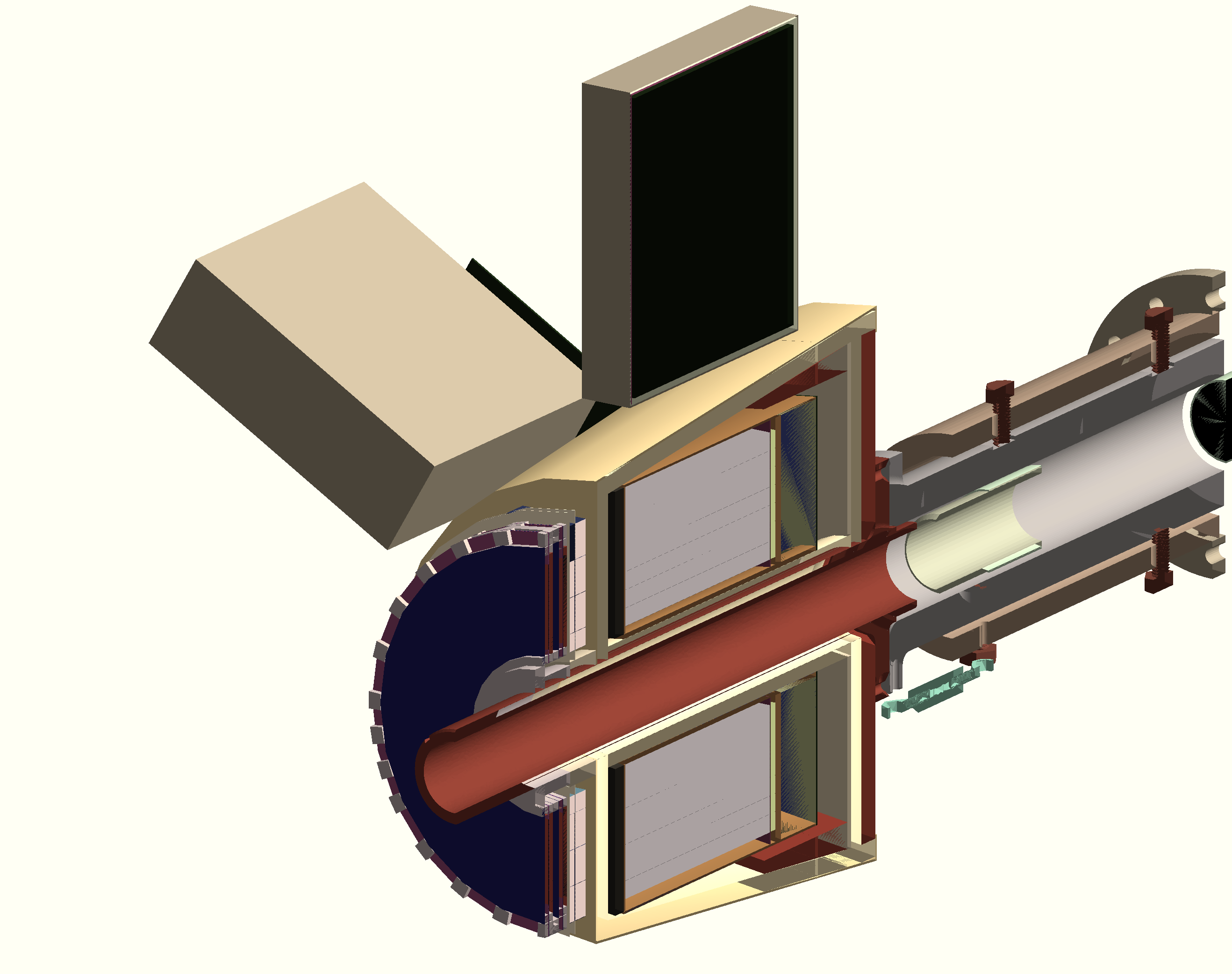}
\caption{Top: the lead target and scattering chamber implementation in GEMC.
	Bottom: details of the implementation of the three Forward Tracker subsystems. As seen by the beam (incident from the left):
	the disks form the FT-Trk; the FT-Hodo scintillators just behind the tracker; the FT-Cal. The hardware mount support.}
\label{fig:ftGeometry}
\end{figure}
The response of FT-Cal crystals is modeled within the simulation as follows. The time window  of the calorimeter is set to $132~\text{ns}$, and all Geant4 steps within the same paddles
and time window are collected in one hit.
The digitization routines are called at the end of each event, after the Geant4 navigation has
propagated all tracks and GEMC has collected all the steps into hits. The deposited energy is converted first to the charge produced at the end of the electronics chain
composed by an avalanche photodiode (APD) and preamplifier, and then to an ADC. The first conversion
is based on the measured charge for cosmic rays that deposit a known energy in the crystals,
while the second one is based on the FADC conversion factor. A smearing on the final
ADC values is added, accounting for the Poisson distribution of photoelectrons produced by
the photosensor, the Gaussian noise of the photosensor and of the preamplifier.
All parameters---the number of photoelectrons per\mev of energy deposited, and the RMS width of the APD noise and of the preamplifier input noise---have been tuned to the experimental data.

\begin{figure}[htbp]
    \centering
    \includegraphics[width=.45\textwidth]{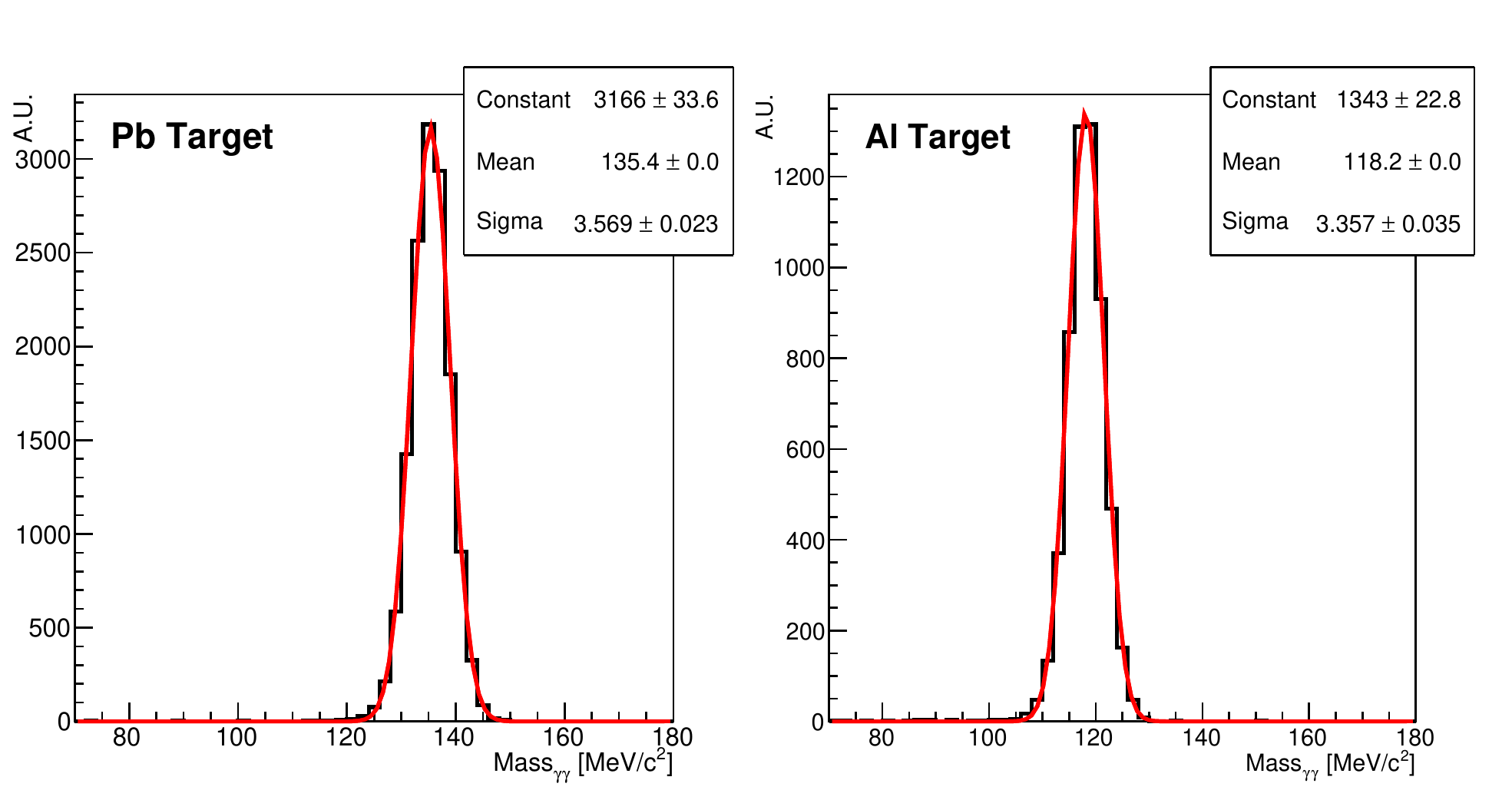}
    \caption{Left: $\gamma\gamma$-invariant mass spectrum of $\pi^0$ produced by the interaction of the beam with Pb target. Right: $\gamma\gamma$-invariant mass spectrum of $\pi^0$ produced by the Al target. In both cases, the mean value and width of the distribution was determined via Gaussian fit (red line).}
    \label{fig:mass_simul}
\end{figure}

The $\pi^{0}\to\gamma\gamma$ events were generated in a fiducial region $2^{\circ}< \theta < 6^{\circ}$ and in a photon energy range $2 < E_{\gamma} < 10\gev$ at Pb target position accordingly with the theoretical distribution and then were passed to GEMC using the LUND data format which encodes the particle IDs, vertices and momenta for each event.\\
The output of the GEMC simulations was analyzed by applying the same clustering algorithms used for experimental data.  
The simulation shows that the overall efficiency is $1.4\%$, providing an expected  $\pi^{0}$-event yield of 1800 $\pm$ 200 for $N_{e} =1.8 \times 10^{14}$.\footnote{This value corresponds to the accumulated charge during Hall-B tests.}  Figure \ref{fig:mass_simul}-Left shows the reconstructed invariant mass peak centered at the expected $\pi^0$ mass and $3.6\mev$ width, corresponding to an energy resolution of 2.9$\%$. \\
Pions produced by the interaction of the beam/photons with the two downstream aluminum foils were also simulated, and the corresponding expected yield was found to be $420\pm100$. The reconstructed $\pi^0$ mass distribution, showed in Fig.~\ref{fig:mass_simul}-Right, has a lower mean value, as a result of the assumption that the vertex was located at the lead target position when calculating the invariant mass. The resulting $\pi^0$ mass peaks associated to Pb and Al production, are separated by  $\sim17\mev$. 

\subsubsection{Data analysis and results}
\label{sec_data_analysis and results}
The offline data reconstruction is performed by applying the same full suite of reconstruction algorithms used in the online analysis, which are implemented in the JANA2 framework and described in Sec.~\ref{subsec:jana}.  
The main task is reconstructing the clusters from the raw information associated with the particles detected in the calorimeter. 
This raw information includes the charge and time for each channel of the detector with a signal above the front-end threshold. 
The first step in this reconstruction pipeline is to determine the energy and time of the individual crystal hits.
The recorded charge of each hit is converted into energy by applying the proper calibration constant, which is determined from a calibration run performed in standard triggered mode. Details on the calibration procedure are reported in~\cite{ACKER2020163475}. 
It is worth noting that the calibration run was performed several weeks before the SRO tests, and that significant radiation damage induced on the crystals was not accounted for.

\begin{figure}[h!]
\centering
\includegraphics[width=0.45\textwidth]{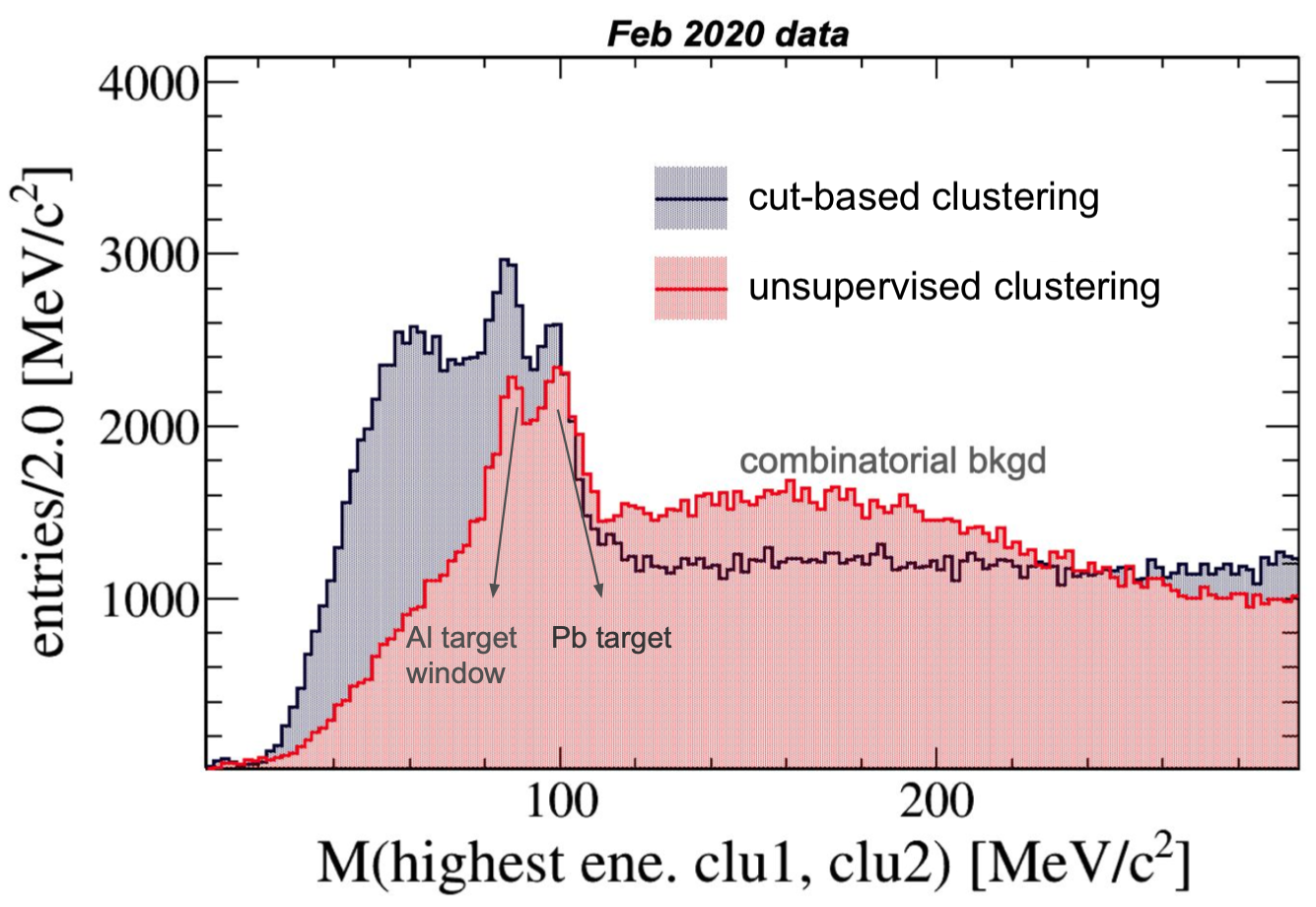}
\caption{Diphoton invariant mass spectrum from the two highest energy clusters obtained with the unsupervised clustering (red) as compared to the standard clustering (black) algorithms.
\label{fig:uncalibrated}
}
\end{figure}

As mentioned in Sec. \ref{subsec:AI_SRO}, one of the nice features of an unsupervised clustering algorithm like HDBSCAN (compared to standard clustering algorithms) is the ability to cluster hits without using cuts on time, spatial or energy information, but rather looking at their correlations. Therefore, the effect of energy miscalibrations does not impact the HDBSCAN clustering performance, but only affects the reconstructed cluster energy. 
Also, because AI-based algorithms such as HDBSCAN are capable of performing with large hit multiplicities and in presence of substantial noise, they are particularly desirable in high luminosity experiments that will operate in SRO mode. Fig.~\ref{fig:uncalibrated} shows a comparison of the diphoton invariant mass spectrum obtained with raw data (uncalibrated), utilizing both the standard and the HDBSCAN clustering algorithms.  
Both spectra are obtained by applying loose fiducial cuts on the reconstructed events, a minimum cluster size of 3 hits, and a minimum cluster energy (dominated by a threshold on hit energies). 

\begin{figure}[htbp]
    \centering
    \includegraphics[width=.5\textwidth]{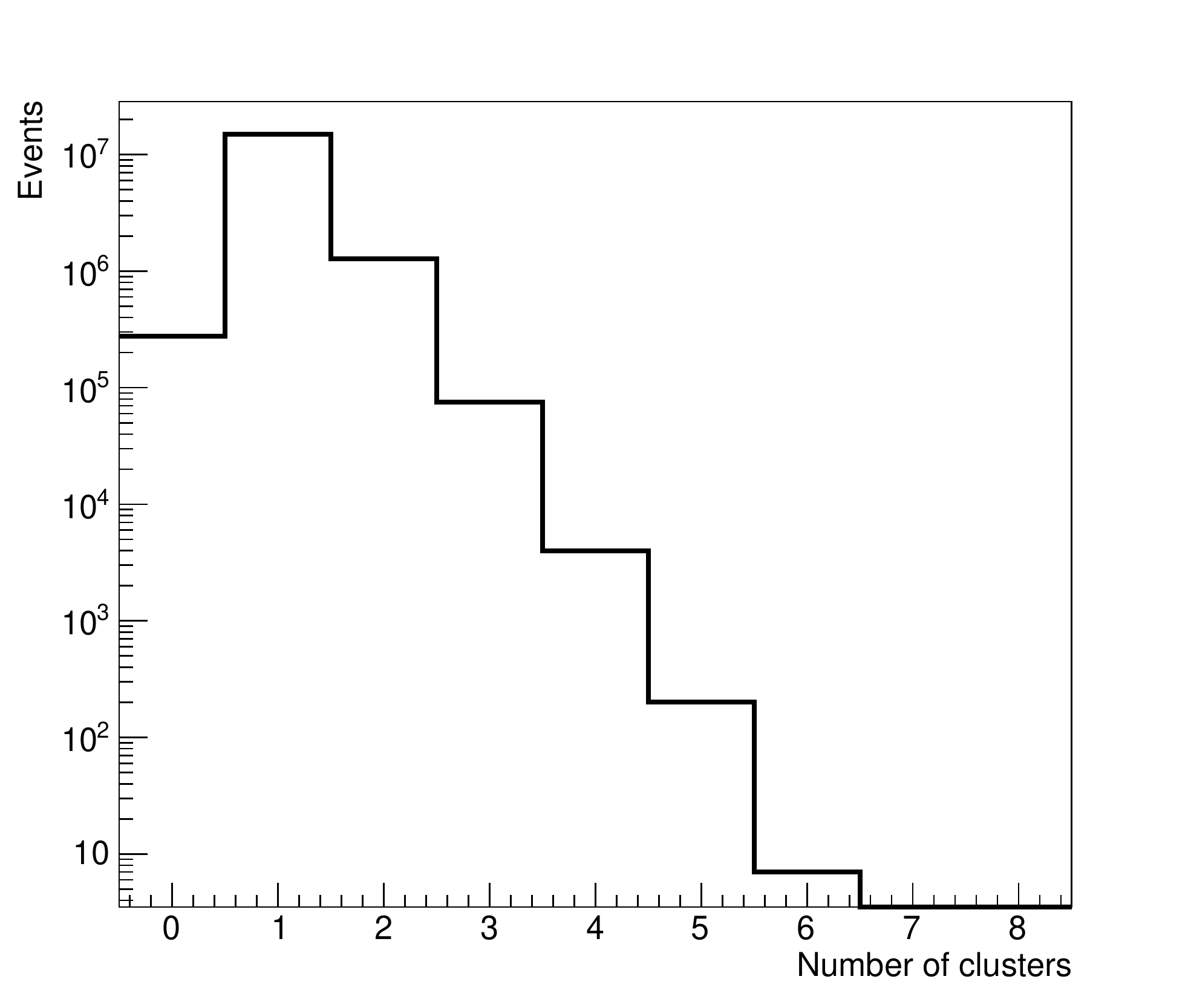}
    \caption{Distribution of number of clusters per event.}
    \label{fig:multi_cluster}
\end{figure}

Remarkably, the unsupervised clustering provided a $\pi^{0}$ yield consistent with the cut-based standard clustering, but with significantly improved signal-to-background ratio in the region sensitive to the $\pi^{0}$ peak.
The standard clustering algorithm utilizes cuts to aggregate hits around the seed of a cluster, and suffers when dealing with data that is not properly calibrated. 
 In the near future we plan to further characterize the performance of this unsupervised approach: preliminary results during our tests have shown a longer runtime of $\sim$30\% relative to the standard clustering algorithm when run in a single thread.   
 This clustering technique seems a promising alternative to traditional cut-based approaches, particularly when deployed online on data that is not fully calibrated, and is capable of rejecting noise hits and dealing with complicated topologies of clusters and higher multiplicities.  This comes at the cost of introducing the 2 hyperparameters described in Sec.~\ref{subsec:AI_SRO}, which can be tuned with dedicated studies. 
 These results show that the AI-based algorithms are robust and provide an efficient event selection on raw data where miscalibrations are expected.\\
In the following, we describe the data analysis performed using the standard clustering algorithm in more detail. 
After correcting for the {\it time walk} effect, the reconstructed hits were sorted by energy. A cluster was identified starting from the seed and adding adjacent crystals with a signal above $50\mev$ and within a time window of $10~\text{ns}$. Once a cluster was identified, the  energy, time and position were computed according to the procedure described in Ref.~\cite{ZIEGLER2020163472}. 
Fig.~\ref{fig:multi_cluster} shows the distribution of the number of clusters per event: the largest bucket corresponds to one cluster, and only about $8\%$ of the events have two or more.
The analysis aimed at identifying the $\pi^{0} \rightarrow \gamma\gamma$ decays only considers two-cluster events.
It uses following selection cuts:
\begin{itemize}
\item two clusters appear within a time window of $10~\text{ns}$ from each other;
\item both clusters have energy above $500\mev$;
\item the sum of the two clusters' energy is less than $8\gev$;
\item the number of crystals involved in each cluster is greater than 3;
\item the opening angle between the two clusters is greater than $2^{\circ}$;
\item the polar angle of both clusters falls within $2.5^{\circ}<\theta<4.5^{\circ}$.
\end{itemize}

\begin{figure}[htbp]
    \centering
    \includegraphics[width=.5\textwidth]{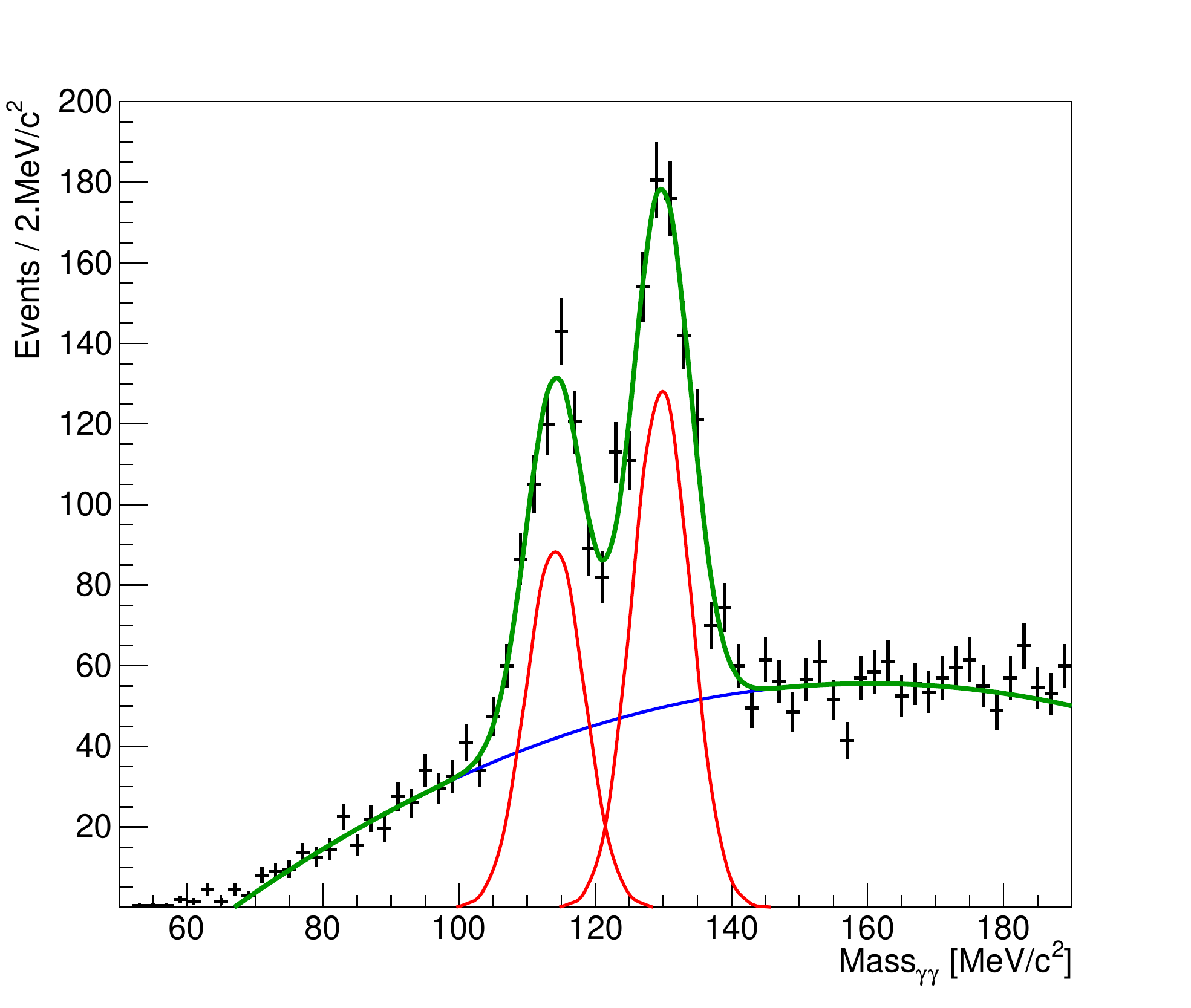}
    \caption{Distribution of $\gamma\gamma$ invariant mass. The two peaks were fit with Gaussian functions (red dashed lines) plus a quadratic polynomial function for the background (blue line). The green line represents the overall fit. 
    As discussed in Fig.~\ref{fig:mass_simul}, the lower mass peak corresponds to the Al window, the larger mass peak to the Pb target.}
    \label{fig:invMass}
\end{figure}

Fig.~\ref{fig:invMass} shows the reconstructed $\gamma\gamma$-invariant mass spectrum. It is characterized by two peaks centered at $114.1\pm2.7\mev$ and $129.6 \pm 2.1\mev$ respectively. 
The presence of the two peaks and the relative distance are in agreement with the  Monte Carlo simulation results reported  in Sec.~\Ref{subsubsec:simulations}. 
We interpreted the higher-mass peak as being due to $\pi^{0}$ production from the lead target, and the other peak as being due to $\pi^{0}$-production from downstream materials (e.g. aluminum windows). The two peaks are centered at lower mass values, relative to the expected values, due to the aforementioned miscalibration.
To determine the $\pi^{0}$ production yield, the $\gamma\gamma$-invariant mass spectrum was fit with two Gaussian functions plus a quadratic polynomial function for the background. The number of events in the first and second peak, determined by integrating the respective Gaussian function from $-3$ to $3\sigma$, were found to be $966 \pm 164$ and $1378 \pm275$, respectively. The latter is in agreement within 30$\%$ of the theoretical expectation for generated $\pi^0$ by the interaction of the beam with the lead target via real and virtual photoproduction mechanisms. The former exceeds the expected yield related to the production from the two Al windows by a factor $\sim4$. This discrepancy could be due to the presence of other materials placed near the two Al windows (e.g. glue, mechanical support) contributing to the $\pi^0$ production and consequently increasing the measured yield.

\subsubsection{JLAB SRO-DAQ performance}

During the Run-2 tests, a study of SRO DAQ performance was conducted. From the front-end, a data rate of about $800~\text{MB}/\text{s}$ per uplink was measured with no data frame dropping (100\% livetime). Since the setup consisted of 3 VXS crates with 6 fiber uplinks, the total data rate reached up to $4~\text{GB}/\text{s}$.\\
To study the performance of the back-end, the front-end thresholds and TriDAS parameters (i.e. the number of instances of HMs and TCPUs) were varied. During tests, the memory occupancy and the CPU load per TriDAS process were checked against the data throughput. An uneven distribution of data sources was found to have a significant impact on TriDAS performance. This is not a surprise, since the system was originally designed for a neutrino telescope, where all detection elements produce almost the same data throughput, providing a well distributed and balanced load to the HM.
The best performance was achieved with a single  memory assignment to fulfill the requirements of every instantiated HM. However, throughput homogeneity is not guaranteed in CLAS12 streams. 
The topology of the physics events created sizeable gradients in the throughput across different sectors of the FT-CAL and FT-HODO detectors. 
The first version of the TriDAS implementation, which is not yet optimized, handles this problem by dimensioning all memory buffers according to the maximum size necessary to accommodate the largest data stream. This of course biases the measured memory occupancy.\\
The front-end thresholds were varied to provide a data   throughput ranging from a few tens of\mbps up to almost $100\mbps$. The HM processes were instantiated on one Linux server with 48 cores, $1~\text{GHz}$ each and $64~\text{GB}$ RAM. The number of HM instances per run were raised from 5 HMs, 10 HMs and 20 HMs. The detector was subdivided in 5, 10 and 20 sectors, accordingly. 
The CPU load increased almost linearly with the number of HM instances, 500\%, 850\% and 1600\%, respectively. This is implicit in the multi-threaded design of TriDAS. Meanwhile, the HM memory occupancy remained almost constant at about $12$--$1~\text{GB}$ per run. This is consistent with the  $500~\text{kB}/\text{channel}/\text{timeslice}$ buffer size, and the fact that the number of HMs is inversely proportional to the number of served channels per HM, which is the total number of FT+Hodo channels, i.e. a constant on the order of $\sim500$.\\  
Ten instances of TCPUs,  each capable of handling 5 timeslices at time,  run on two CPU servers. As mentioned in Sec.~\ref{TRIDAS_sec}, the TCPU implements different trigger-level algorithms. The Level~1 performance was found to be strongly affected by hit sorting in the considered timeslice. The profiling of this  nonlinear performance  was reported in~\cite{Chiarusi_2017}. The Level~2 trigger was not always used, in order to determine the impact of running TriDAS with or without the JANA algorithms. 
The CPU load per TCPU instance ranged from 400\%  without any JANA trigger, to 800\% including the {\em standard clustering},  the $1:10$ scaler and the minimum bias selection algorithms, and, up to 1600\% when processing  the AI clustering. Generally the memory usage remained within $20$--$24~\text{GB}$. However, it doubled when running the AI algorithm, indicating the need for optimization.  

\section{Future work}
\subsection{The ERSAP SRO framework}
The Environment for Real-time Streaming, Acquisition and Processing (ERSAP) framework is an effort at JLab to develop streaming readout and data processing systems that will satisfy future experiments at the laboratory which are currently in various stages of planning. The goal is to develop a common framework for building both streaming data acquisition and data stream processing pipelines. The new experiments are eager to take advantage of SRO  technologies, due to the well-known intrinsic limitations of the triggered readout systems. The ERSAP framework organizes a data stream processing application into a network of interconnected actors which communicate via message-passing and execute user-defined algorithms referred to as `engines'. The framework abstracts these engines as micro-services and provides them with a run-time environment. This includes handling the network programming, data serialization, and general IO. ERSAP also provides tools for engine configuration and scaling, potentially freeing the user from having to write multi-threaded code. The only requirement ERSAP imposes is that the user-defined engines must adhere to a simple data-in/data-out interface. 

A key goal for ERSAP is to be flexible enough to allow data processing applications to evolve. It should encourage the implementation of new ideas and technologies while preserving the integrity of existing data pipelines. To this end, it provides three basic components: a reactive Actor micro-service, a communication channel between Actors (which serves as a data-stream pipe or conveyer belt), and finally an application orchestrator. During operation, a stream of data-quanta will flow through the directed graph of reactive microservices, and the network itself will define the higher-level application logic.

One basic difference between ERSAP and other frameworks is that, rather than moving instructions between Actors, the data is moved instead. An incoming data-quantum triggers the execution of an actor (in other words, the actor \emph{reacts} to the incoming message). Communications between actors is restricted to message passing, and the channels between actors are specified externally to the actors themselves. A key consequence of this design is that actors are programmatically independent, i.e. they can be built and deployed as stand-alone processes.

ERSAP is a framework that uses an independent, reactive actor model and a flow-based programming paradigm. It encourages a functional decomposition of the overall data processing application into small single-function artifacts called micro-services. These artifacts should be easy to understand, develop, deploy and debug. Because they are programmatically independent, they can be readily scaled and individually optimized, unlike the components of a monolithic application. Another important advantage is fault tolerance: individual actors are allowed to crash and be automatically restarted without bringing down the entire pipeline.

\subsection {GEMC SRO}
The GEMC simulation framework has been updated to collect hits from multiple events based on the Time at the Readout Electronics (TRE) of each Geant4 step, as provided by the digitization routine. The data is organized and collected in the same data structures used by the VTP, called Frames.

During digitization, the TRE is calculated for each Geant4 step
by taking into account the track navigation in the various material, the response of the detector, and the propagation delays due to the hardware and cables. A dedicated GRunAction class handles:

\begin{itemize}
    \item creating the VTP frame buffers
    \item linking the distribution of hits from various events to frame buffer ids using the TRE
    \item flushing the buffers to disk  
\end{itemize}

The digitized data identifier is converted into hardware addresses (organized by  crate/slot/channel) using the detector translation tables.
Each VTP frame refers to the corresponding hardware crate, uniquely identified by absolute time $t$ and the frame duration
$\Delta t_f$ using the formula $\text{id}=\floor(t)/\Delta t_f$.
The frame data includes:

\begin{itemize}
\item the Frame Payload, which contains the collection of hits, represented as a vector of unsigned int, and also includes the hardware address.
\item the Frame Header, which contains the payload properties and time of frame.
\end{itemize}

The distribution of Geant4 events into frames is summarized in \F{eventCollection}.

\begin{figure}
\centering
\includegraphics[width=0.99\columnwidth,keepaspectratio]{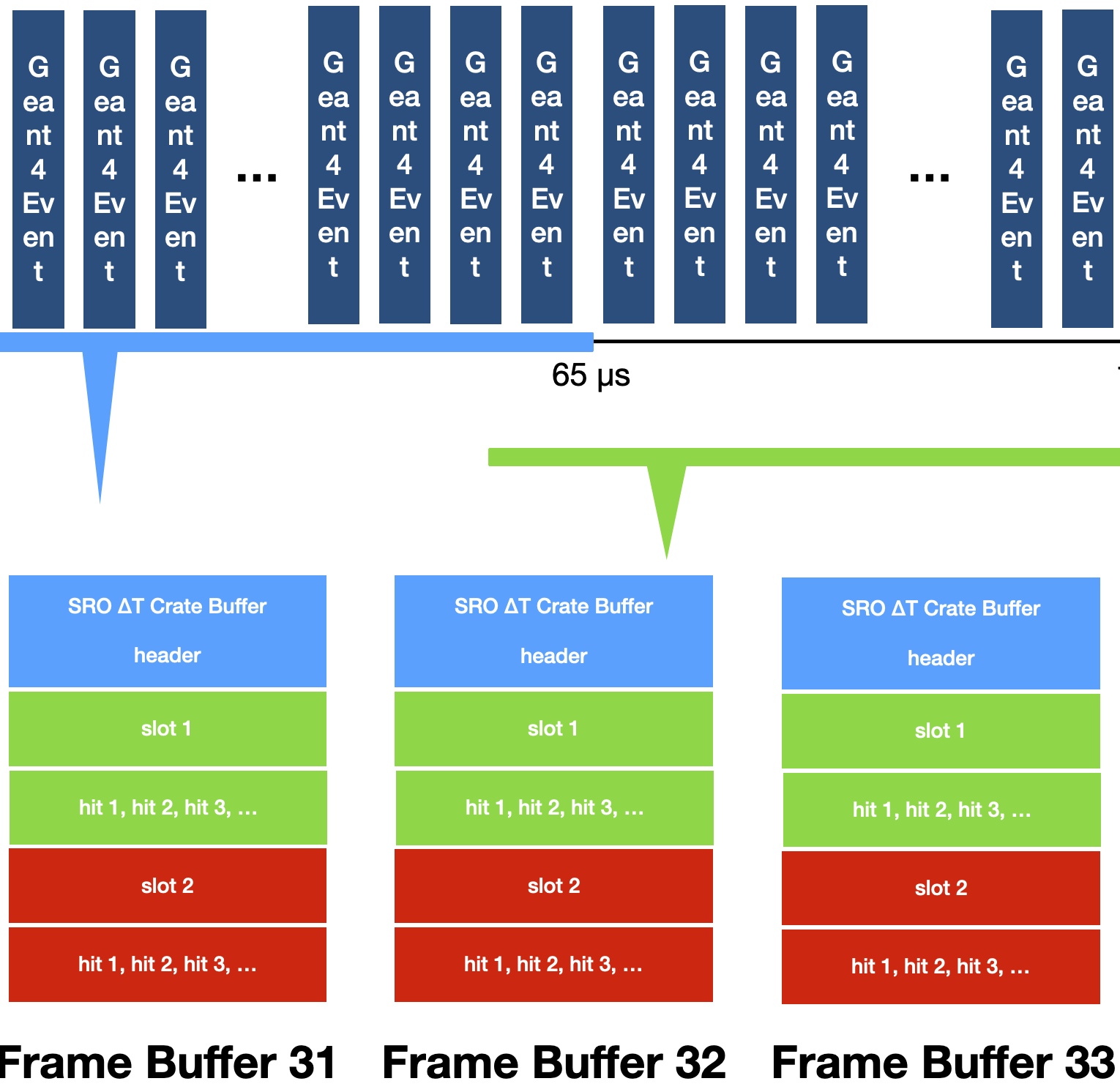}
\caption{Hits from a Geant4 event can end on different frames depending on the digitization time,
propagation time, etc. Many Geant4 events can have hits in the same Frame.
When no more Geant4 event can fill a frame, that frame is written out to disk. }
\label{fig:eventCollection}
\end{figure}
The GEMC output is available in various formats, which are identical in content: text (ASCII),
EVIO (the Jefferson Lab data acquisition format)~\cite{evio}, ROOT and VTPSRO,
the binary format used for the CLAS12 SRO. GEMC was used to produce both event-based output and a frame-based streaming readout output that emulates the JLab SRO front-end.

\section{Summary}

Streaming readout is a powerful and flexible option in data acquisition adopted by many current and future experiments, which would be particularly beneficial for electron machines at the intensity frontier (e.g. JLab and the EIC). Despite its many advantages and the simplicity (in principle) of implementation, the superiority of streaming readout in real applications still needs to be proven. In this paper we reported results of on-beam tests of the first Jefferson Lab implementation of a full SRO DAQ system, including front-end electronics (JLab-FADC250 and INFN-\WVB digitizers), back-end software (TriDAS) and high-level analysis framework (JANA2). Tests were performed at the lab with two different experimental setups. 
In the first, we exposed the EIC PbWO$_4$ crystal EM calorimeter prototype to the Hall-D Pair Spectrometer test beam  for a direct comparison of triggered and SRO performance. Results showed that the SRO performed as expected, providing a calorimeter energy resolution compatible with data collected using a traditional triggered DAQ.
The second setup included the Hall-B CLAS12 Forward Tagger calorimeter and hodoscope to measure a physics channel, inclusive $\pi^0$ hadroproduction, during a standard high-intensity electron-beam production run. Results were compared to the  expected yield calculated with a realistic  theoretical model of the reaction making use of a sophisticated Geant4 simulation to determine the detector efficiency. The good agreement between prediction and measurement provided a significant validation of the SRO DAQ  performance during a realistic electron scattering experiment. Furthermore, the implementation in the framework of AI-supported real-time tagging and selection algorithms, demonstrating how the SRO DAQ provides new capabilities well beyond the standard triggered DAQ. Based on these positive results, the current framework is being upgraded to a microservice architecture (ERSAP) to build a common framework for both streaming readout and offline data processing.

\section*{Acknowledgments}

We would like to acknowledge the CLAS12 and GlueX collaborations as well as the JLab technical staff for their accommodation and support of this effort. 

The INFN Group has been supported by Italian Ministry of Foreign Affairs (MAECI) as Projects of Great Relevance within Italy/US Scientific and Technological Cooperation under grant n.
MAE0065689-PGR00799.

This material is based upon work supported by the U.S. Department of Energy, Office of Science, Office of Nuclear Physics under contract DE-AC05-06OR23177. 
The work of CF is supported by the U.S. Department of Energy, Office of Science, Office of Nuclear Physics under grant No. DE-SC0019999.

\section*{Data Availability Statement}
The data are usually not deposited. Please contact the corresponding author for details

\bibliographystyle{unsrturl}

\bibliography{SRO_EPJ}

\end{document}